\newlength{\intwidth}

\documentclass{jfm}

\usepackage{amssymb}

\usepackage{array}
\usepackage{amsmath}
\usepackage{latexsym}
\usepackage{bezier}
\usepackage{psfrag,graphicx}
\usepackage{bm}
\usepackage{float}
\usepackage{multirow}
\usepackage{mathrsfs}
\usepackage{color}
\usepackage{enumerate}
\usepackage{natbib}
\usepackage{nicefrac}
\usepackage[colorlinks, citecolor=blue]{hyperref}
\usepackage{overpic}
\usepackage{rotating}
\usepackage{verbatim}

\begin{document}

\title[Three-dimensional coherent structures in a curved pipe flow]{Three-dimensional coherent structures in a curved pipe flow}

\author{
  Runjie Song\aff{1}
  \and Kengo Deguchi\aff{1}
}
 
 \affiliation{
   \aff{1}School of Mathematics, Monash University, VIC 3800, Australia
}

\maketitle

\begin{abstract}


Dean's approximation for curved pipe flow, valid under loose coiling and high Reynolds numbers, is extended to study three-dimensional travelling waves. Two distinct types of solutions bifurcate from the Dean’s classic two-vortex solution. The first type arises through a supercritical bifurcation from inviscid linear instability, and the corresponding self-consistent asymptotic structure aligns with the vortex-wave interaction theory. 
The second type emerges from a subcritical bifurcation by curvature-induced instabilities and satisfies the boundary region equations. 
Despite the subcritical nature of the second type of solutions, it is not possible to connect their solution branches to the zero-curvature limit of the pipe. However, by continuing from known self-sustained exact coherent structures in the straight pipe flow problem, another family of three-dimensional travelling waves can be shown to exist across all Dean numbers. The self-sustained solutions also possess the two high-Reynolds-number limits.
While the vortex-wave interaction type of solutions can be computed at large Dean numbers, their branch remains unconnected to the Dean vortex solution branch.
\end{abstract}

\section{Introduction}\label{sec:introduction}

Fluid flow through a curved pipe is encountered in various applications, from engineering equipment to biological systems. It is well-known that unidirectional laminar flow is not achievable in this system, as the balance between centrifugal force and pressure is disrupted, resulting in the formation of steady cross-stream vortices named after Dean's seminal works \citep{Dean_1927,Dean_1928}. 
The earliest qualitative observations of this flow pattern can be traced back to \cite{boussinesq1868memoire}. Over the past century, Dean vortices have remained a fundamental example of secondary flows in fluid dynamics.


Extensive research of curved pipes has been conducted through experiments \citep{White_1929,Ito_1959,Sreenivasan_Strykowski_1983,kuhnen2014,kuhnen2015}, theory \citep{Dean_1927,Dean_1928,Dyke_1978,DL_1989,BM_2014,BM_2017}, and numerical computations \citep{collins_1975,Patankar_Pratap_Spalding_1975,Webster_Humphrey_1997,huttl2001,Piazza_Ciofalo_2011}. Readers seeking quick access to this vast body of work are encouraged to consult review articles by \cite{BTY_1983}, \cite{vashisth2008} and \cite{Vester_2016}. Numerical computations of curved pipe flows can be broadly categorised into those that use the so-called loose-coiling approximation, originating from Dean's work, and those that do not, and this study belongs to the former category. The full curved pipe flow problem involves two parameters: the Reynolds number and the (dimensionless) curvature of the pipe. The loose-coiling approximation is valid when the curvature is small, and the Reynolds number is large, allowing the behavior of Dean vortices to be captured by a single parameter known as the Dean number. Historically, this approximation was particularly valuable when computational resources were limited. When the Dean number is small, perturbation theory can be applied, providing a useful check on numerical computations for finite Dean numbers.  

In the 1980s and 1990s, significant interest centered on the non-uniqueness of steady vortex structures in the loose-coiling approximation system. Two families of four-vortex solutions were numerically discovered \citep{Benjamin_1978,nandakumar1982,Winters_1987,Yanase_1989,DL_1989},  and are later elegantly reconstructed via perturbation expansion \citep{BM_2014,BM_2017}.
These flow states can be regarded as what are now called exact coherent structures.
Identifying exact coherent structures has become one of the major focuses in shear flow research, guided by dynamical systems theory (see \cite{Kerswell_2005}, \cite{Eckhardt2007}, for example). However, previous studies on Dean vortices have only addressed two-dimensional stationary structures that are invariant along the pipe's axial direction. The primary goal of this research is therefore to extend these theoretical and numerical results to three-dimensional travelling wave-type exact coherent structures, providing a broader theoretical understanding of the complex dynamics in curved pipe flows.

Although stable travelling wave states have been observed under specific parameters in experiments and numerical simulations \citep{Webster_Humphrey_1993, Webster_Humphrey_1997}, to the best of the authors' knowledge, systematic continuation study of corresponding exact coherent structures by Newton's method has not been conducted yet. 
Stability analysis of the Dean vortex with respect to three-dimensional perturbations serves as an obvious first step towards obtaining nonlinear travelling waves by bifurcation analysis. However, somewhat surprisingly, such stability analysis was not pursued until the recent work by \cite{Canton_Schlatter_orlu_2016}.

Another pathway to finding a three-dimensional travelling wave is through homotopy continuation from the exact coherent structures obtained in a straight pipe flow \citep{faisst2003,wedin2004,PK07,pringle2009}. In the absence of curvature, no linear instability arises in the laminar Hagen-Poiseuille flow, and thus the transition to turbulence is necessarily triggered by finite-amplitude perturbations. Exact coherent structures are crucial for an understanding of such subcritical transition problems \citep{Kerswell_2005,Eckhardt2007}. The physical mechanism by which coherent structures are maintained, independent of laminar flow instabilities, is commonly explained by a cyclic interaction between the rolls, streaks and waves (self-sustaining process, see \cite{hamilton1995}, \cite{waleffe1997}). The cycle naturally emerges from the large-Reynolds-number asymptotic expansion of exact coherent structures in the high-Reynolds-number limit, known as the vortex-wave interaction theory (VWI, see \cite{hall1991}, \cite{HALL_SHERWIN_2010}, \cite{deguchi2014}, \cite{ozcakir2016}).


At medium to high curvatures of the pipe, \cite{Sreenivasan_Strykowski_1983}, \cite{Webster_Humphrey_1993,Webster_Humphrey_1997} and \cite{kuhnen2014} identified a supercritical transition  characterised by the emergence of a travelling wave. However, when the pipe curvature is small, the neutral curve recedes to higher Reynolds numbers, making subcritical transition, as observed in straight pipe flow problems, more dominant \citep{kuhnen2015}. 
Expanding on these findings, \cite{canton2020} conducted detailed investigation around the pipe curvature at which the nature of the transition changes from subcritical to supercritical. They observed that within a narrow range of pipe curvatures, the flow can exhibit both sustained turbulence and a stable travelling wave, with two competing attractors in the phase space. Moreover, the supercritical Hopf bifurcation point for the three-dimensional travelling waves perfectly aligns with the neutral curve computed by the full Navier-Stokes approach \citep{Canton_Schlatter_orlu_2016}. 

One of the remaining unanswered questions is how the three-dimensional travelling waves that emerge in the supercritical transitions relate to the exact coherent structures self-sustained at the straight pipe case. This inquiry is closely tied to the works by \cite{Nagata_1988,Nagata_1990}, where exact coherent structures in plane Couette flow were found through continuation from secondary flows induced by system rotation. Another key question in this paper is the relationship between the Dean vortices in the loose-coiling approximation and the VWI theory, both of which apply in the high-Reynolds-number limit. Bridging these theories seems promising, given that the VWI originated from the asymptotic theory for G\"ortler vortices \citep{hall_smith_1988}.

For plane Couette flow, \cite{Deguchi_Hall_Walton_2013} showed that when the streamwise wavelength of exact coherent structures is comparable to the Reynolds number, the VWI approximation breaks down and must be replaced by boundary region equations (BRE). More recently, \cite{dokoza2023} considered the same limit to explain the large-scale coherent structures observed in direct numerical simulations in \cite{lee2019}. We shall show that two high-Reynolds-number limits, VWI and BRE, are also possible for the curved pipe flow problem.

The rest of the paper is organised as follows. In the next section, we shall formulate the curved pipe flow problem using Navier-Stokes equations. The loose-coiling approximation and its extensions to the three-dimensional travelling waves will be discussed. Section~\ref{sec:BF_2v} first studies the large Reynolds number asymptotic properties of the stability of the Dean vortices with respect to three-dimensional perturbations. In the same section, we shall also study the bifurcation of nonlinear travelling waves.
Section~\ref{sec:Conti_pipe} is devoted to the continuation of exact coherent structures from the straight pipe problem. Finally, in section~\ref{sec:conclusion}, we present our conclusions and discuss the implications of the results.



\section{Formulation of the problem}\label{sec:2}

\begin{figure}
\centering
\begin{overpic}[width=0.9 \textwidth]{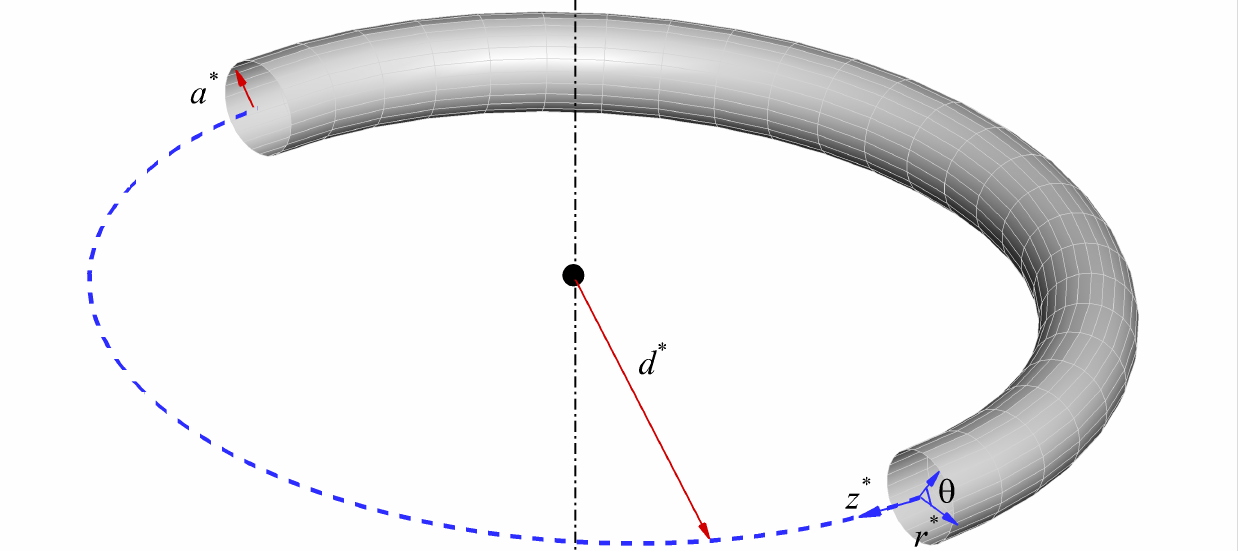}
\end{overpic}
\caption{A sketch of the curved pipe studied in this paper. The gray surface represents a section of a torus with minor and major radii denoted by 
$a^*$ and $d^*$, respectively.
The flow field is described by the orthogonal coordinates $(r^*,\theta,z^*)$. 
}
\label{fig:sketch_model}
\end{figure}
Consider an incompressible Newtonian viscous fluid with the density $\rho^*$ and the dynamic viscosity $\mu^*$ flowing through a curved circular pipe. As sketched in figure~\ref{fig:sketch_model}, we denote the radius of the curvature of the pipe centreline as $d^*$ and the radius of the pipe as $a^*$, with the latter chosen as the length scale. 
Following \cite{Germano_1982,Germano_1989}, the orthogonal coordinates $(r^*,\theta,z^*)$ are installed to describe the radial, circumferential and streamwise directions, respectively. 
We assume that the flow is driven by a constant pressure gradient $G^*$. In the absence of the curvature, the centreline velocity of the laminar Hagen-Poiseuille flow is given by $U_c^*=\frac{G^* a^{*2}}{4 \mu^*}$, and we adopt this as the velocity scale. 
The non-dimensional velocity field $(u,v,w)$ and pressure $p$ are governed by the Navier-Stokes equations
\begin{subequations}\label{fulleq}
\begin{eqnarray}
\omega \frac{\partial w}{\partial z}+\frac{\partial u}{\partial r}+\frac{u}{r}+\frac{1}{r}\frac{\partial v}{\partial \theta}+\kappa \omega \{u\cos \theta-v\sin \theta\}=0,~~~~~~\\
\frac{Du}{Dt}-\frac{v^2}{r}-\kappa \omega w^2\cos \theta=-\frac{\partial p}{\partial r}+\frac{1}{R} \left \{\left (\frac{1}{r}\frac{\partial}{\partial \theta}-\kappa \omega \sin \theta \right )S_1-\omega\frac{\partial S_3}{\partial z} \right \},~~~~~~\\
\frac{Dv}{Dt}+\frac{uv}{r}+\kappa \omega w^2 \sin \theta=-\frac{1}{r}\frac{\partial p}{\partial \theta}+\frac{1}{R}\left \{\omega \frac{\partial S_2}{\partial z}-\left (\frac{\partial}{\partial r}+\kappa \omega \cos \theta \right)S_1\right \},~~~~~~\\
\frac{Dw}{Dt}+ \kappa \omega w(u\cos \theta -v\sin \theta)=-\omega \frac{\partial p}{\partial z}+\frac{1}{R} \left \{4+\left (\frac{1}{r}+\frac{\partial}{\partial r}\right)S_3-\frac{1}{r}\frac{\partial S_2}{\partial \theta}\right \},~~~~~~
\label{eq:NS_non}
\end{eqnarray}
where $r=r^*/a^*$, $z=z^*/a^*$ and
\begin{eqnarray}
\frac{D}{Dt}=\frac{\partial}{\partial t}+\omega w \frac{\partial }{\partial z}+u\frac{\partial }{\partial r}+\frac{v}{r}\frac{\partial }{\partial \theta},\qquad \omega=\frac{1}{1+\kappa r \cos \theta},\\
S_1=\frac{1}{r}\frac{\partial u}{\partial \theta}-\frac{\partial v}{\partial r}-\frac{v}{r},~~
S_2=\omega \frac{\partial v}{\partial z}+\kappa \omega w \sin \theta-\frac{1}{r}\frac{\partial w}{\partial \theta},\\
S_3=\frac{\partial w}{\partial r}+\kappa \omega w\cos \theta-\omega\frac{\partial u}{\partial z}.
\end{eqnarray}
\end{subequations}
The flow has two parameters, the Reynolds number and the non-dimensional curvature:
\begin{equation}
    R=\frac{U_c^* a^*\rho^*}{\mu^*}, \qquad \kappa= \frac{a^*}{d^*}. 
\end{equation}
The no-slip conditions $u=v=w=0$ are imposed at $r=1$. In the streamwise direction, the flow is assumed to be periodic with a period $\frac{2\pi}{\alpha}$, where $\alpha$ is the wavenumber. Our Reynolds number $R$ is based on the pressure gradient; therefore, the bulk velocity $Q$ (i.e., normalised flux) is one of the appropriate quantities to diagnose the flow state. 

As \cite{Dean_1927} realised, the combined parameter $K\equiv 2\kappa R^2$ plays an important role when the curvature $\kappa$ is small. 
This is one of the widely used definitions of the `Dean number' found in the literature. 
In experiments, however, the flux is easier to control, and $De\equiv RQ\kappa^{1/2}=Q(K/2)^{1/2}$ is more commonly used (see \cite{Vester_2016}). Since $Q$ depends on $R$ in a non-trivial manner, numerical computations are necessary to link $K$ with $De$. 

If the curvature $\kappa$ is not small, the flow cannot be controlled by the Dean number alone (see e.g. \cite{topakoglu1967}), and our study does not cover such a parameter regime.


\subsection{Dean vortices} \label{sec:dean_v}
Suppose the flow is steady and does not depend on $z$.
The loose coiling approximation corresponds to the asymptotic limit of $R \to \infty$, $\kappa \to 0$ while keeping the Dean number $K=2\kappa R^2$ as an $O(1)$ quantity. 
Substituting the asymptotic expansions
\begin{subequations}\label{eq:long_UVW}
\begin{eqnarray}
u=R^{-1}U(r,\theta)+\cdots,\qquad
v=R^{-1}V(r,\theta)+\cdots,\\
w=W(r,\theta)+\cdots, \qquad
p=R^{-2}P(r,\theta)+\cdots,
\end{eqnarray}
\end{subequations}
into (\ref{fulleq}) and retaining the leading order terms, we obtain the set of equations
\begin{subequations}\label{deaneq}
\begin{eqnarray}
\frac{\partial U}{\partial r}+\frac{U}{r}+\frac{1}{r}\frac{\partial V}{\partial \theta}=0,~~~~~\\
U\frac{\partial U}{\partial r}+\frac{V}{r}\frac{\partial U}{\partial \theta}-\frac{V^2}{r}-\frac{K}{2} W^2\cos \theta=-\frac{\partial P}{\partial r}+
\triangle_2 U-\frac{U}{r^2}-\frac{2}{r^2}\frac{\partial V}{\partial \theta} ,~~~~~\label{deaneqb}\\
U\frac{\partial V }{\partial r}+\frac{V}{r}\frac{\partial V}{\partial \theta}+\frac{UV}{r}+\frac{K}{2} W^2 \sin \theta=-\frac{1}{r}\frac{\partial P}{\partial \theta}+
\triangle_2 V-\frac{V}{r^2}+\frac{2}{r^2}\frac{\partial U}{\partial \theta}
,~~~~~\label{deaneqc}\\
U\frac{\partial W}{\partial r}+\frac{V}{r}\frac{\partial W}{\partial \theta}=4+
\triangle_2 W.~~~~~
\end{eqnarray}
\end{subequations}
Here $\triangle_2=\partial_r^2+r^{-1}\partial_r+r^{-2}\partial_{\theta}^2$ and the no-slip conditions $U=V=W=0$ must be fulfilled at $r=1$. The equations (\ref{deaneq}) are equivalent to (15)-(18) of \cite{Dean_1928}. 


\begin{figure}
\centering
\begin{overpic}[width=0.96 \textwidth]{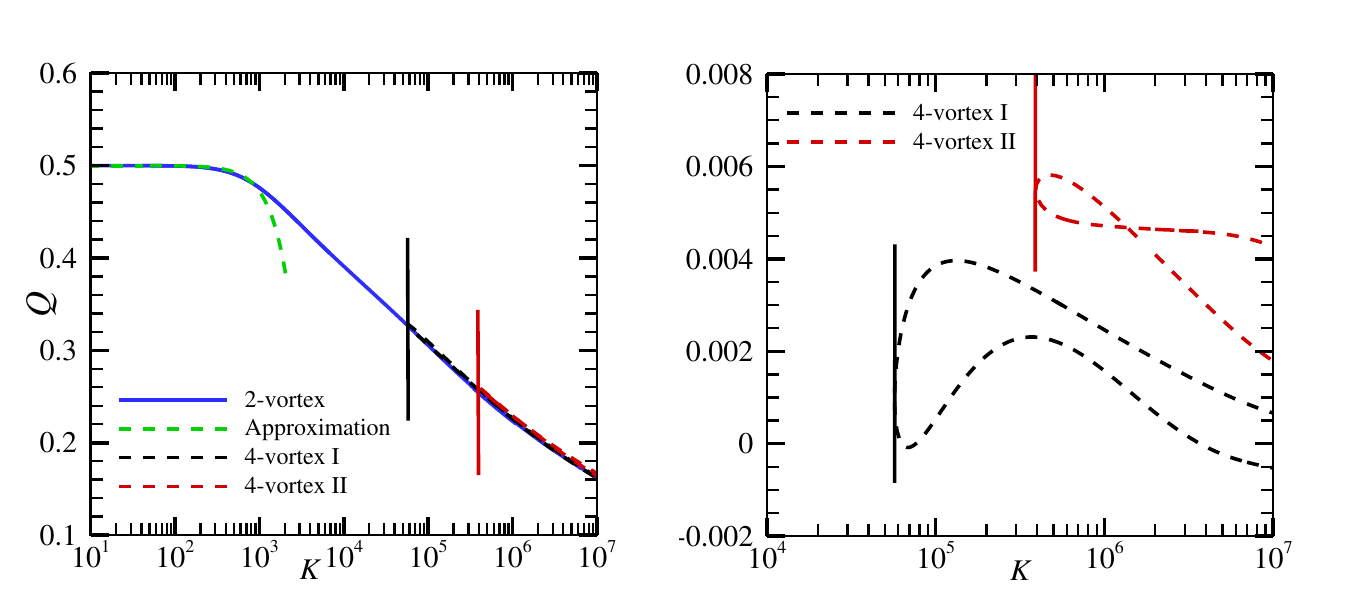}
\put(-0.5,39.3){(a)}
\put(47,39.3){(b)}
\put(47,21){\begin{turn}{90}{$\Delta Q$}\end{turn}}
\put(62,25){$K_1$}
\put(72,28){$K_2$}
\put(26,25){$K_1$}
\put(31,12){$K_2$}
\end{overpic}
\caption{
(a): Dependence on $K$ of the total average velocity $Q$. 
The dashed green curve is the approximation (\ref{eq:appde}), neglecting the $O(K^6)$ terms. (b) The same 4-vortex solutions as in (a), but expressed in terms of the deviation of $Q$ from the 2-vortex solution.
The values of $K$ at the saddle-node bifurcation points are $K_1\approx 5.71 \times 10^4$, $K_2 \approx 3.89\times 10^5$. 
} 
\label{fig:wb_2v_4v}
\end{figure}
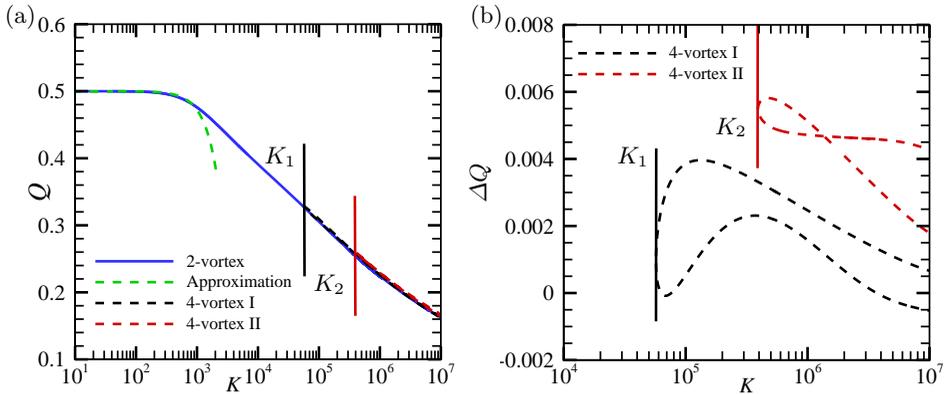

When $K=0$, a unidirectional laminar flow solution $(U,V,W)=(0,0,1-r^2)$ exists. As the curvature increases, a pair of streamwise vortices develops; hereafter this state is referred to as the 2-vortex solution. \cite{Dean_1928}  found this solution using perturbation approach, and obtained the following approximation for the bulk velocity $Q=\frac{1}{\pi}\int^{2\pi}_0\int^1_0 W (r,\theta) r dr d\theta$:
 \begin{eqnarray}\label{eq:appde}
Q_2 \approx \frac{1}{2}\left (1-0.0306\left (\frac{K}{576} \right)^2+0.0120\left (\frac{K}{576} \right )^4+O(K^6) \right ).
\end{eqnarray}
Here the subscript 2 represents this result is valid for the 2-vortex solution. 
Here, $K=576$ corresponds to $De\approx 16.59$. 
Attempts to extend the radius of convergence of the perturbation expansion were made by \cite{Dyke_1978}, and more recently
\cite{BM_2014,BM_2017} successfully reproduced the two families of 4-vortex solutions previously reported numerically (\cite{Benjamin_1978}, \cite{Winters_1987}, \cite{Yanase_1989} , \cite{DL_1989}). Figure~\ref{fig:wb_2v_4v}-(a) summarises the variation of $Q$ for 2- and 4-vortex solutions. To gain a clearer understanding of the bifurcations, it is helpful to summarise the results in terms of the deviation from the 2-vortex value, $\Delta Q\equiv Q-Q_2$ (panel b).
The 2-vortex solution is known to be stable for $z$-independent perturbations in the range of $K$ shown in the figure. 
 However, it becomes unstable against more general perturbations at a critical $K$, as we will see in section \ref{sec:BF_2v}. 

\subsection{Numerical methods} \label{sec:numerics}
Our aim is to extend the above argument to three-dimensional travelling waves. 
An examination reveals that, except for the terms involving $K$, (\ref{deaneq}) matches the Navier-Stokes equations in cylindrical coordinates, but with a unit Reynolds number and no $z$-dependence. Therefore, a naive approach would be to simplify the full governing equations (\ref{fulleq}) to
\begin{subequations} \label{eq:Dean_effect}
\begin{eqnarray}
 \frac{\partial w}{\partial z}+\frac{\partial u}{\partial r}+\frac{u}{r}+\frac{1}{r}\frac{\partial v}{\partial \theta}=0,~~~\\
\frac{Du}{Dt}-r^{-1}v^2- \frac{K}{2} \left (\frac{w}{R} \right)^2 \cos \theta = -\frac{\partial p}{\partial r}+ \frac{1}{R}(\triangle u-r^{-2}u-2r^{-2}\frac{\partial v}{\partial \theta}),~~~ \label{eq:D1} \\
\frac{Dv}{Dt}+r^{-1}uv+ \frac{K}{2} \left (\frac{w}{R} \right)^2 \sin \theta=-r^{-1}\frac{\partial p}{\partial \theta}+\frac{1}{R}(\triangle v -r^{-2}v+2r^{-2}\frac{\partial u}{\partial \theta}),~~~\label{eq:D2} \\
\frac{Dw}{Dt}=-\frac{\partial p}{\partial z}+\frac{1}{R}(4+\triangle w),~~~  \label{eq:D3}
\end{eqnarray}
\end{subequations}
where $\frac{D}{Dt}=\partial_t+u\partial_r+r^{-1}v\partial_{\theta}+w\partial_z$ and $\triangle=\partial_r^2+r^{-1}\partial_r+r^{-2}\partial_{\theta}^2+\partial_z^2$. 
It is easy to verify that (i) when $K=0$, the equations (\ref{eq:Dean_effect}) become the three-dimensional Navier-Stokes equations governing the straight pipe flow problem, and (ii) the leading order part of the Dean vortex solutions satisfy the equations (\ref{eq:Dean_effect}).

The use of the above equations can be justified by asymptotic analysis. To clarify the discussion, we shall define a terminology: we will refer to a reduced system as Asymptotic Preserving Reduction (APR) when it contains all the essential components to yield the leading order solution of the full equations (\ref{fulleq}). We shall demonstrate that, in the same limit considered by \cite{Dean_1927}, there are two possible sets of reduced equations for three-dimensional coherent structures, and that (\ref{eq:Dean_effect}) is APR of both of them.


All numerical results in this paper are based on (\ref{eq:Dean_effect}), including figure \ref{fig:wb_2v_4v}. In order to find nonlinear travelling wave solutions $\mathbf{u}(r,\theta,z-ct)$, we apply a Galilean shift to eliminate the time dependence. 
Our numerical code is based on \cite{DEGUCHI_NAGATA_2011}, where the poloidal-toroidal decomposition $\mathbf{u}= \mathcal{W}(r)\mathbf{e}_z+\nabla \times \nabla \times (\phi(r,\theta,z) \mathbf{e}_r)+\nabla \times (\psi(r,\theta,z) \mathbf{e}_r)$ is used. The continuity is automatically satisfied, and the independent equations can be obtained by operating $\mathbf{e}_r\cdot \nabla \times  \nabla \times$,  $\mathbf{e}_r\cdot \nabla \times$, and the $\theta$-$z$ average to the momentum equations. The basis functions for the poloidal potential $\phi$, toroldal potential $\psi$, and the mean flow $\mathcal{W}$ are the same as those used in \cite{Deguchi_Walton_2013}. A Fourier-Galerkin method is used in the $\theta$ and $z$ directions, while a Chebyshev collocation method is employed in the $r$ direction. This transforms the problem into a set of algebraic equations, with the spectral coefficients and phase speed $c$ as unknowns, which can then be solved using Newton's method. The truncation level of the expansions is specified using the triplet $(L,M,N)$, where $L$ is the degree of Chebyshev polynomials, and $M$ and $N$ are the order of Fourier series in the $\theta-$ and $z-$ directions, respectively. 

Since we have the Jacobian matrix at hand, stability analysis can be readily performed. The complete list of eigenvalues is first computed by the LAPACK routine ZGGEV and the unstable modes are tracked in the parameter space by the well-known Rayleigh quotient iteration scheme. 

\section{Bifurcation from the 2-vortex solution} \label{sec:BF_2v}

\subsection{Linear stability of the 2-vortex solution} \label{sec:LST_2v}

\begin{figure}
\centering
\begin{overpic}[width=0.96 \textwidth]{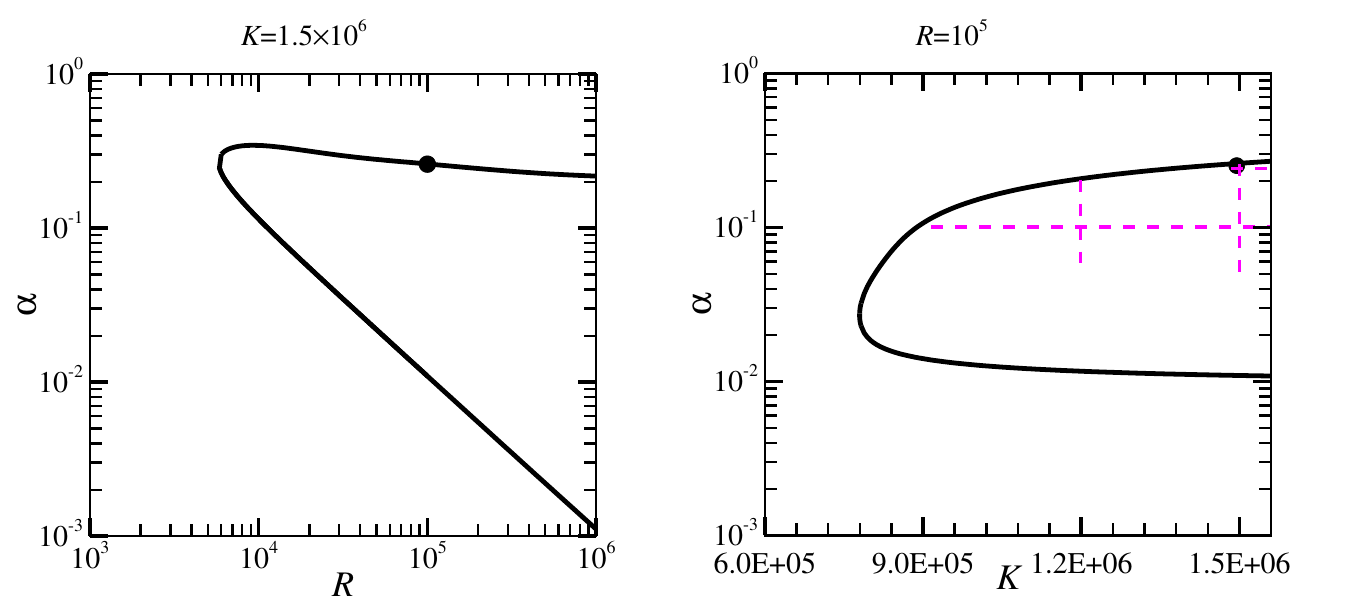}
\put(-2,39.3){(a)}
\put(47,39.3){(b)}
\end{overpic}
\caption{
The stability of the 2-vortex solution found by the Orr-Sommerfeld equations (\ref{eq:OS}). 
(a) The neutral curve in the $ \alpha$-$R$ plane at $K=1.5 \times 10^6$. The upper curve is the inviscid mode. (b) The neutral curve in the $\alpha$-$K$ plane at $R=10^{5}$. The bullets represent the same point in the parameter space. The magenta dashed lines are parameter range studied in figure \ref{fig:Dn=750000_change_zBn_R=100000_VWI}. The spatial resolution is checked using up to $(L,M)=(50,50)$. 
}
\label{fig:neutral_VWI}
\end{figure}

The linear stability of the Dean vortices can be analysed by introducing a perturbation:
$\mathbf{u}=(R^{-1}U,R^{-1}V,W)+ (\tilde u, \tilde v, \tilde w)$, $p=P+\tilde{p}$. 
The perturbation is assumed to be proportional to an infinitesimally small amplitude $\delta >0$ and  a normal mode as
\begin{eqnarray}\label{waveexp}
(\tilde u, \tilde v, \tilde w, \tilde p)=\delta (\hat {u}(r,\theta),\hat {v}(r,\theta),\hat {w}(r,\theta),\hat {p}(r,\theta))e^{{\rm i}\alpha z+\sigma t}+\text{c.c.},
\end{eqnarray}
where c.c. stands for the complex conjugate. The real part of the complex growth rate $\sigma= \sigma_r + {\rm i} \sigma_i $ determines the stability.

Given that Dean's limit corresponds to the high Reynolds number regime, the base flow may be dominated by 
$W$. Moreover, the advection effect by that component is much stronger than the curvature effects of $O(R^{-2})$ (see (\ref{eq:Dean_effect})).
If we are allowed to neglect $U$, $V$, and the terms proportional to $K$, the stability can be found by the Orr-Sommerfeld equation generalised for the base flow varying in two directions:
\begin{subequations}\label{eq:OS}
\begin{eqnarray}
\frac{\partial \hat{u}}{\partial r}+\frac{\hat{u}}{r}+\frac{1}{r}\frac{\partial \hat{v}}{\partial \theta}+ {\rm i} \alpha \hat{ w}=0,\\ 
(\sigma+{\rm {i}}\alpha W) \hat{u}=-\frac{\partial \hat {p}}{\partial r}+\frac{1}{R}\left \{(\triangle_2 -\alpha^2)  \hat{u}  -\frac{\hat{u}}{r^2}-\frac{2}{r^2}\frac{\partial \hat{v}}{\partial \theta}\right \},\\
(\sigma+ {\rm {i}}\alpha W) \hat{v}=-\frac{1}{r}\frac{\partial \hat{p}}{\partial \theta}+\frac{1}{R}\left \{(\triangle_2 -\alpha^2) \hat{v}-\frac{\hat{v}}{r^2}+\frac{2}{r^2}\frac{\partial \hat{u}}{\partial \theta}\right \},\\
(\sigma+{\rm {i}}\alpha W) \hat{w}+\hat{u}\frac{\partial W}{\partial r}+\frac{\hat{v}}{r}\frac{\partial W}{\partial \theta}=- {\rm i} \alpha { \hat{p}}+\frac{1}{R}(\triangle_2 -\alpha^2) \hat{w}.
\end{eqnarray}
\end{subequations}
The no-slip conditions $\hat{u}=\hat{v}=\hat{w}=0$ are imposed at $r=1$.
Using the 2-vortex solution as the base state at $K=1.5\times 10^6$, the eigenvalue problem yields the neutral curve shown in figure~\ref{fig:neutral_VWI}-(a). The upper neutral curve tends to a constant value of $\alpha$ as $R$ increases, which is a typical feature of inviscid instability. Note that for neutral modes the viscous terms are still important around the critical layer, where $W-c$ vanishes. Nevertheless, for sufficiently large $R$, the generalised Orr-Sommerfeld result matches with the inviscid result (\cite{Deguchi_2019}). Since computations become challenging at very high Reynolds numbers, this paper will use $R=10^5$ to infer results involving inviscid waves. Figure~\ref{fig:neutral_VWI}-(b) shows the neutral curve obtained by varying $K$ at that fixed $R$. The instability only exists for $K \gtrsim 7.78\times 10^5$ ($De\gtrsim$ 289.69).
The inviscid mode serves a starting point of the analysis of the VWI type nonlinear solutions in section \ref{sec:BF_VWI_2v}.

\begin{figure}
\centering
\begin{overpic}[width=0.96 \textwidth]{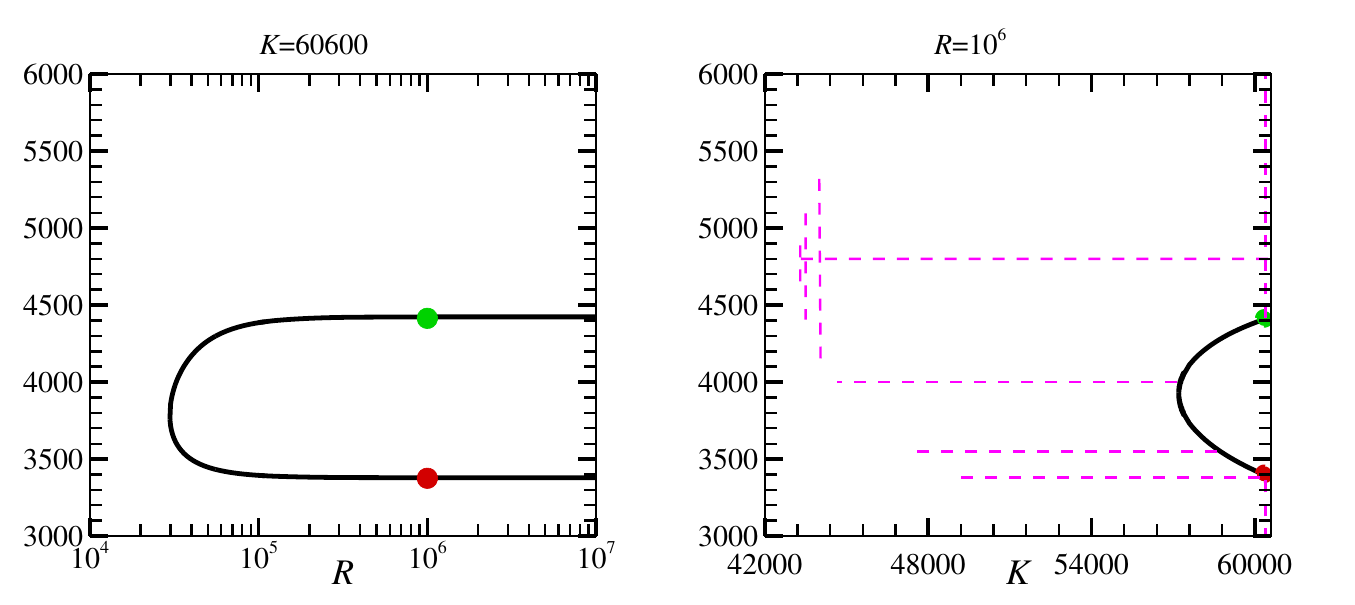}
\put(-2,39.3){(a)}
\put(-2,21){$ \alpha_0$}
\put(47,39.3){(b)}
\end{overpic}
\caption{
The stability results based on the approximated equation  (\ref{eq:OS}) linearised around the 2-vortex solution.
(a) The neutral curve of the curvature mode in the $\alpha_0$-$R$ plane at $K=60600$. The long-wavelength limit (\ref{eq:long}) is achieved as $R\rightarrow \infty.$
(b) The neutral curve in the $ \alpha_0$-$K$ plane at $R=10^{5}$. 
The bullets represent the same point in the parameter space. The magenta dashed lines are the parameter range studied in figure \ref{fig:change_zDn_different_alphaR_upper}.
Resolution is checked using up to $(L,M)=(50,50)$. 
}
\label{fig:neutral}
\end{figure}
Along the lower neutral curve in figure~\ref{fig:neutral_VWI}-(a), the wavenumber $\alpha$ behaves like $O(R^{-1})$, which is a typical signature of the emergence of the long-wavelength mode. This observation motivates us to take 
the limit $R\rightarrow \infty$ while keeping $\alpha_0\equiv R\alpha$ as a constant, similar to the method used for unidirectional parallel flows \citep{Smith_1979_2,Cowley_Smith_1985}.
However, in this limit, the advection effect due to $W$ becomes $W\partial_z=O(R^{-1})$, making the advection effects due to $U$ and $V$ non-negligible. Furthermore, from the continuity equation, $\tilde{u}$, $\tilde{v}$ are smaller than $\tilde{w}$ by a factor of $O(R^{-1})$, similar to Dean's argument, which necessitates retaining the curvature terms. Formally, the long-wavelength limit can be obtained by rescaling $\alpha=R^{-1}\alpha_0$, $\sigma=R^{-1}{\sigma_0}$ and writing 
\begin{eqnarray}
(\tilde u, \tilde v, \tilde w, \tilde p)=\delta R^{-1} (\hat {u}(r,\theta),\hat {v}(r,\theta),R \hat {w}(r,\theta),R^{-1}\hat {p}(r,\theta))e^{R^{-1}({\rm i}{\alpha_0} z+{\sigma_0} t)}+\text{c.c.},~~~~
\end{eqnarray}
in (\ref{eq:Dean_effect}). The leading order problem can be found as 
 \begin{subequations}\label{eq:long}
\begin{align}
& \frac{\partial \widehat{u}}{\partial r}+\frac{\widehat{u}}{r}+\frac{1}{r}\frac{\partial \widehat{v}}{\partial \theta}+ {\rm i} 
 \alpha_0 \widehat{ w}=0,\\ 
& ( \mathcal{L} +\frac{ \partial U }{\partial r}  + \frac{1}{r^2} ) \widehat{u}+(\frac{1}{r}\frac{\partial U}{\partial \theta} - \frac{2V}{r} +\frac{2}{r^2}\frac{ \partial }{ \partial \theta } ) \widehat{v} -K W \widehat{w}\cos \theta+\frac{ \partial \widehat{ p}}{\partial r}=0, \\
& ( \mathcal{L} +\frac{1}{r}\frac{ \partial V }{ \partial \theta}  + \frac{1}{r^2} +\frac{U}{r} ) \widehat{v}+(\frac{1}{r}\frac{\partial (r V)}{  \partial r}  -\frac{2}{r^2}\frac{ \partial }{ \partial \theta } ) \widehat{u} +K W \widehat{w}\sin \theta+\frac{1}{r}\frac{ \partial \widehat{ p}}{ \partial \theta}=0, \\
&\mathcal{L} \widehat{w}+\frac{\partial W}{\partial r} \widehat{ u} +\frac{1}{r}\frac{\partial W}{ \partial \theta} \widehat{v}=0.
\end{align}
 \end{subequations}
Here we have defined the operator $\mathcal{L} = ({\sigma_0} + {\rm i}  \alpha_0 W +U \frac{\partial}{\partial r}+ \frac{V}{r}\frac{\partial}{\partial \theta}-\triangle_2 )$ to simplify the equations. The usual no-slip conditions complete the eigenvalue problem. 

Figure~\ref{fig:neutral}-(a) shows the stability of the 2-vortex solution by using the linearised version of (\ref{eq:Dean_effect}). Along both branches of the neutral curve, as $R\rightarrow \infty$, the value of ${\alpha_0}$ tends to a constant, which corresponds the limit shown in (\ref{eq:long}). Interestingly, this result suggests that our analysis detects a new mode, distinct from the long-wavelength mode observed in figure~\ref{fig:neutral_VWI}-(a). Hereafter the latter mode is referred to as the `curvature mode' as its existence depends on the presence of the terms proportional to $K$.

The solid curve in figure~\ref{fig:neutral}-(b) shows the neutral curve obtained with $R=10^6$, which is sufficiently large to observe the converged limiting solution.
This figure clearly shows that the curvature mode exists only when $K$ is larger than the critical value $5.72\times 10^4$. In terms of the flux based parameter, this critical point corresponds to $De\approx 110$. It is noteworthy that recently \cite{LCR_2024} studied the stability of the 2-vortex solution to long-wavelength, three-dimensional perturbations using full Navier-Stokes equations (\ref{fulleq}). Their critical Dean number, $De \approx 113$, observed around the loose coiling limit parameter regime, is well compared with our results.


\subsection{Bifurcation from the inviscid mode: VWI} \label{sec:BF_VWI_2v}

From the neutral points obtained above, bifurcations of nonlinear travelling wave solutions are anticipated. We denote the phase speed as $c=\sigma/\alpha$. Of course, $c$ must be purely real for travelling waves. 

Here, we focus on bifurcations of VWI type solutions from the inviscid mode computed in figure~\ref{fig:neutral_VWI}. When the amplitude of the wave-like perturbation reaches a certain level, it begins to affect the Dean vortices through the Reynolds stress. Among these stress terms, the important ones are those that appear in the momentum equations in the $r-$ and $\theta-$ directions, as the velocities in these components are smaller than in the streamwise direction.
Therefore, the Dean equations (\ref{deaneq}) and the Orr-Sommerfeld equations (\ref{eq:OS}) may be coupled via the extra terms $F_r$ and $F_{\theta}$ to the left-hand sides of (\ref{deaneqb}) and (\ref{deaneqc}), respectively, where
\begin{subequations}\label{Rstress}
\begin{eqnarray}
F_r= \frac{R^2\delta^2}{r}\left \{\frac{\partial (r\hat{u}\hat{u}^*)}{\partial r}+\frac{\partial (\hat{u}\hat{v}^*)}{\partial \theta}-\hat{v}\hat{v}^*\right ]+\text{c.c.},\\
F_{\theta}=\frac{R^2\delta^2}{r}\left \{
\frac{\partial (r\hat{u}\hat{v}^*)}{\partial r}+\frac{\partial (\hat{v}\hat{v}^*)}{\partial \theta}+\hat{u}\hat{v}^*
\right \}+\text{c.c.},
\end{eqnarray}
\end{subequations}
and the asterisks denote the complex conjugation.
This combined system is similar to the viscous regularised version of the vortex-wave interaction system used in \cite{Blackburn_Hall_Sherwin_2013} for plane Couette flow. 

The regularised VWI system still depends on $R$. To find the appropriate large Reynolds number asymptotic limit the approach of \cite{HALL_SHERWIN_2010} must be used. In light of (\ref{Rstress}), one might consider balancing the Reynolds stress in Dean's equations with $\delta=R^{-1}$, but this is not correct. 
The reason is that at $r=r_c(\theta)$, where $W-c$ vanishes, the inviscid approximation of (\ref{eq:OS}) breaks down. This necessitates the introduction of a critical layer of thickness $R^{-1/3}$ around $r=r_c(\theta)$. Outside of that layer, the correct leading order part of the asymptotic expansion is given by (\ref{eq:long_UVW}) for the $z$-independent `vortex' part, and by (\ref{waveexp}) with $\delta=R^{-7/6}$ for the wave part. This peculiar exponent arises from the matching of solutions inside and outside the critical layer.

Within the critical layer, slightly different asymptotic expansions must be used, and careful analysis, similar to \cite{HALL_SHERWIN_2010}, reveals that the vortex components are subject to the jump conditions
\begin{subequations}\label{jumps}
\begin{eqnarray}
\frac{r_c'}{r_c}\left [\frac{\partial V}{\partial r}\right ]^{r_c+}_{r_c-}
=\left [\frac{\partial U}{\partial r}\right ]^{r_c+}_{r_c-}
=\frac{r_c'}{r_c}J_1(\theta),\qquad [P ]^{r_c+}_{r_c-}=J_2(\theta),
\end{eqnarray}
where 
\begin{eqnarray}
J_1(\theta)=\frac{C}{\gamma^{5/3}B^5r_c^3}\left \{ \left (-\frac{7}{2}\frac{B'}{B}-\frac{5}{3}\frac{\gamma'}{\gamma}-2\frac{r_c'}{r_c}\right ) \left |\frac{\partial \hat{p}}{\partial \theta}\right |^2 +\frac{\partial}{\partial \theta}\left |\frac{\partial \hat{p}}{\partial \theta}\right |^2\right \},\\
J_2(\theta)=\frac{C}{\gamma^{5/3}B^5r_c^3}\left (2B-1-\frac{r_c''}{r_c} \right )\left |\frac{\partial \hat{p}}{\partial \theta}\right |^2,
\end{eqnarray}
\end{subequations}
$B(\theta)=1+(r_c'/r_c)^2$, $\gamma(\theta)=\frac{\alpha}{B} \frac{\partial W}{\partial r}|_{r=r_c}$, and
$C=2\pi (2/3)^{2/3}(-2/3)!\approx 12.8454$. 
The primes denote the ordinary differentiation. Those jump conditions play the same physical role as the Reynolds stress terms $F_r$ and $F_{\theta}$.
The fully reduced VWI closure is therefore  (\ref{deaneq}), (\ref{jumps}), and the inviscid version of (\ref{eq:OS}). In principle, this system can be obtained by substituting the asymptotic expansions into the full equations (\ref{fulleq}) and performing some straightforward algebraic manipulations. 



The regularised VWI system is an APR of the fully reduced VWI, and the latter system is easier to solve as we do not need to impose the jump conditions explicitly. The terms appearing in that system are a subset of those in (\ref{eq:Dean_effect}). The nonlinear solutions of the regularised VWI system can therefore be obtained by omitting the computations of unnecessary terms in the numerical code described in section \ref{sec:numerics}. It should also be remarked that the equations obtained by linearising the regularised VWI system around the 2-vortex solution corresponds exactly to the eigenvalue problem used to compute figure~\ref{fig:neutral_VWI}. Therefore, by using the eigenvector of the neutral solution as the initial value for the Newton method, a nonlinear travelling wave solution can be obtained around the neutral curve. 

\begin{figure}
\centering
\begin{overpic}[width=0.96 \textwidth]{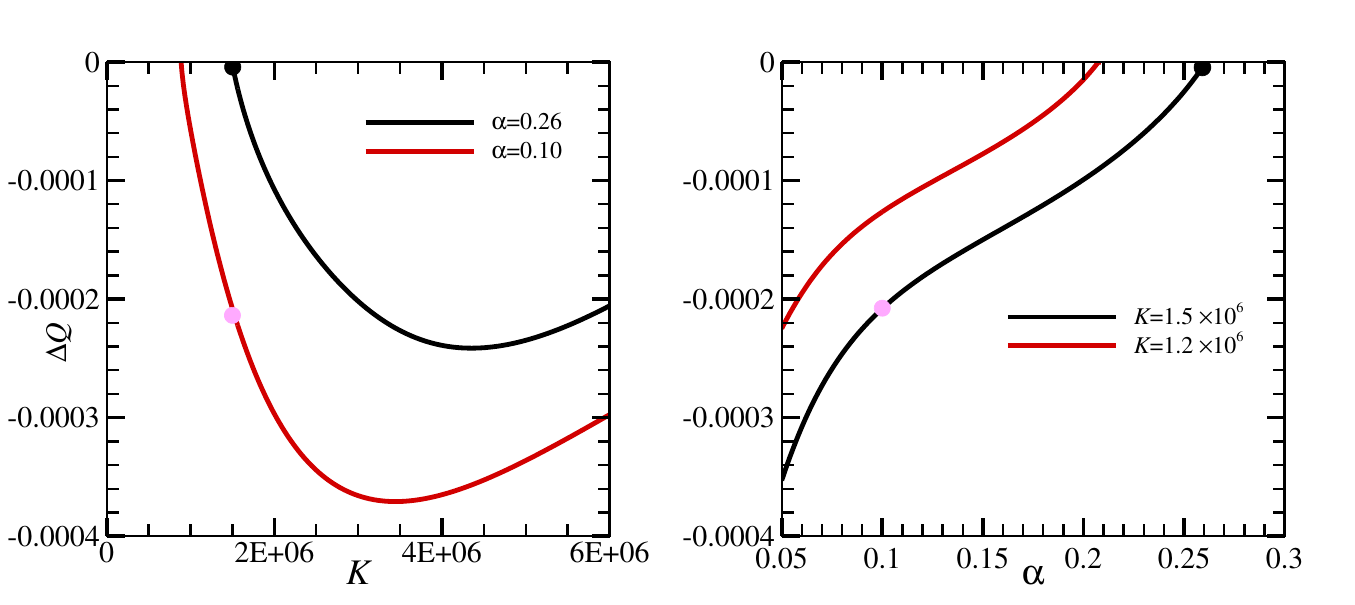}
\put(-2,39.3){(a)}
\put(47,39.3){(b)}
\end{overpic}
\caption{
Bifurcations of the VWI type travelling wave solutions from the inviscid mode. The regularised VWI system with $R=10^5$ is used for computation. The bifurcation point indicated by the black bullet corresponds to the same point shown in figure~\ref{fig:neutral_VWI}. (a) The results for fixed wavenumbers.
(b) The results for fixed Dean numbers. Resolution is checked using up to $(L,M)=(70,50)$. Note that in the regularised VWI, no harmonics are involved in the $z$ direction.  }
\label{fig:Dn=750000_change_zBn_R=100000_VWI}
\end{figure}
\begin{figure}
\centering
\begin{overpic}[width=0.99 \textwidth]{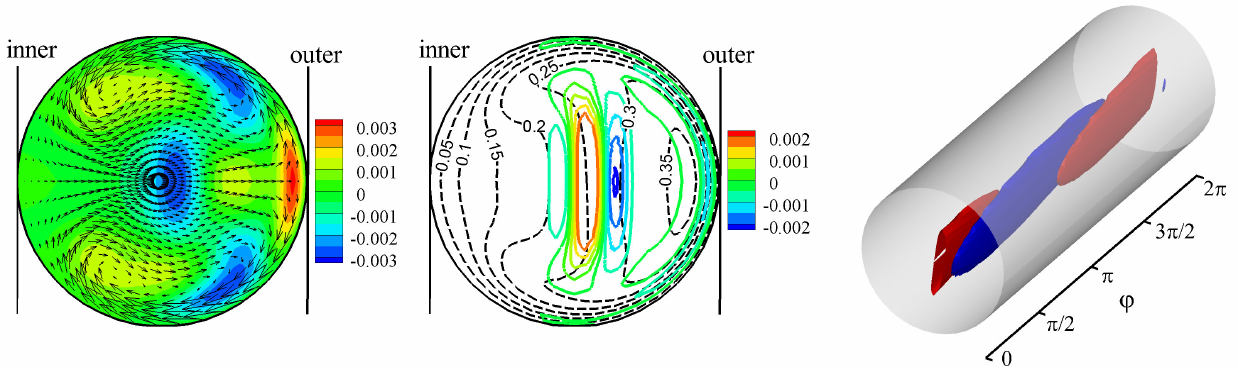}
\put(-4,24.5){(a)}
\put(30,24.5){(b)}
\put(66,24.5){(c)}
\end{overpic}
\caption{The flow structure of the VWI type solution at $(K ,R, \alpha) =(1.5 \times 10^6, 10^5, 0.1)$, corresponding to the pink point in figure~\ref{fig:Dn=750000_change_zBn_R=100000_VWI}. The phase speed is $c\approx 0.2597$.
(a) The vector field represents the roll velocities $\overline{u}$ and $\overline{v}$. The colour indicates the deviation of the streak velocity $\overline{w}$ from that of the 2-vortex solution at the same $K$. 
(b) The black dashed curves represent the isocontours of $\overline{w}$, while the coloured curves show the isocontours of $\tilde{\omega}_z$ at $\varphi=0$. 
(c) The red/blue surface depicts the positive/negative isosurfaces of $\tilde{\omega}_z$ at a magnitude of $0.002$. The phase is defined by $\varphi = \alpha (z - c t)$. } 
\label{fig:VWI_vis}
\end{figure}

The bifurcation diagram is obtained as figure~\ref{fig:Dn=750000_change_zBn_R=100000_VWI}. The parameter range computed in the two panels corresponds to the magenta dashed lines in figure~\ref{fig:neutral_VWI}-(b), indicating that the bifurcation is supercritical. Figure~\ref{fig:VWI_vis}  shows the flow structure of the travelling wave solution at the pink 
bullet point indicated in figure~\ref{fig:Dn=750000_change_zBn_R=100000_VWI}. 
Here and hereafter in order to visualise the flow field we adopt the flow decomposition
$\mathbf{u}=(\overline{u},\overline{w},\overline{w})+ (\tilde u, \tilde v, \tilde w)$ with the first term on the right hand side represents the $z$-averaged part.
More specifically, we apply this decomposition for the leading order part of the asymptotic expansions. Thus, for VWI, $(\overline{u},\overline{v},\overline{w})=(R^{-1}U,R^{-1}V,W)$ and the wave components can be used with their notation unchanged.
The arrows in panel (a) indicate that the components $\overline{u}$ and $\overline{v}$, traditionally called the `roll' component in the VWI and self-sustaining process theory, inherit the large scale swirls by the 2-vortex solution. The black contours in panel (b) show the `streak' component, $W$. The colour map in panel (a) illustrates the extent to which the component $\overline{w}$ deviates from that of the 2-vortex solution.
Panel (c) shows the three-dimensional structure of the `wave' component visualised by the isosurfaces of the streamwise vorticity $\tilde{\omega}_z=r^{-1}(\frac{\partial (r\tilde{v})}{\partial r}-\frac{\partial \tilde{u}}{\partial \theta})$. 
Here, we switched the streamwise variable to the phase  $\varphi = \alpha (z - c t) \in [0,2\pi] $. As seen in panel (b), the wave structure is concentrated around the critical level, at which $\overline{u}$ matches the phase speed $c\approx 0.2597$. This amplification, also seen in other numerical works \citep{wang2007,viswanath2009,McKEON_SHARMA_2010}, is precisely due to the fact that inviscid neutral waves have singularity there. It can be easily shown that $\tilde{\omega}_z$ is $O(R^{-7/6})$ outside the critical layer, while inside it scales as $O(R^{-1/2})$.
The flow field satisfies
\begin{equation}\label{shiftre2}
    [u,v,w](r,\theta,\varphi)=[u,-v,w](r,-\theta+\pi,\varphi+\pi).
\end{equation}
This equation implies that the flow field remains unchanged when shifted by half a period in the streamwise direction and reflected about the $\theta=0,\pi$ axis. This symmetry is referred to as the sinuous mode in the context of secondary flow instability in boundary layer flows \citep{Hall_Horseman_1991,Yu_Liu_1994}.



\subsection{Bifurcation from the curvature mode: BRE}\label{sec:BF_BRE_2v}

A similar bifurcation analysis can be performed for the curvature mode shown in figure~\ref{fig:neutral}. Recall that in Section \ref{sec:LST_2v}, we rescaled the wavenumber and the growth rate as $\alpha = R^{-1}{\alpha_0}$ and $\sigma = R^{-1}{\sigma_0}$. This motivates us to employ the expansions
\begin{subequations}\label{eq:long_UVW2}
\begin{eqnarray}
u=R^{-1}U(r,\theta,X,T)+\cdots,\qquad
v=R^{-1}V(r,\theta,X,T)+\cdots,\\
w=W(r,\theta,X,T)+\cdots, \qquad
p=R^{-2}P(r,\theta,X,T)+\cdots,
\end{eqnarray}
\end{subequations}
using $X=R^{-1}z$, $T=R^{-1}t$. Substituting these into the full Navier-Stokes equations (\ref{fulleq}) yields the reduced problem
\begin{subequations}\label{eq:BRE}
\begin{eqnarray}
\frac{\partial W}{\partial X}+\frac{\partial U}{\partial r}+\frac{U}{r}+\frac{1}{r}\frac{\partial V}{\partial \theta}=0,\\
\mathcal{D}U-\frac{V^2}{r}-\frac{K}{2} W^2\cos \theta=-\frac{\partial P}{\partial r}+
\triangle_2 U-\frac{U}{r^2}-\frac{2}{r^2}\frac{\partial V}{\partial \theta},\\
\mathcal{D}V+\frac{UV}{r}+\frac{K}{2} W^2 \sin \theta=-\frac{1}{r}\frac{\partial P}{\partial \theta}+
\triangle_2 V-\frac{V}{r^2}+\frac{2}{r^2}\frac{\partial U}{\partial \theta},\\
\mathcal{D}W=4+
\triangle_2 W,
\end{eqnarray}
\end{subequations}
correct to $O(R^{-2})$ in the Dean limit. Here we have defined the operator
\begin{eqnarray}
\mathcal{D}=\frac{\partial }{\partial T}+ W \frac{\partial }{\partial X} +U\frac{\partial }{\partial r}+\frac{V}{r}\frac{\partial }{\partial \theta}
\end{eqnarray}
and impose the boundary conditions $U=V=W=0$ at $r=1$.
The equations (\ref{eq:BRE}) have similar structure to the nonlinear equations for the G\"ortler vortex problem formulated by \cite{Hall_1988}, and (4.2) in  \cite{Smith_1976}. This type of equations is more commonly referred to as boundary region equations (BRE) for the study of boundary layer flows (see the discussion in \cite{WU_ZHAO_LUO_2011} and \cite{Deguchi_Hall_Walton_2013}), and we will adopt this terminology. One can easily confirm that the reduced problem (\ref{eq:Dean_effect}) is an APR of BRE.

\begin{figure}
\centering
\begin{overpic}[width=0.96 \textwidth]{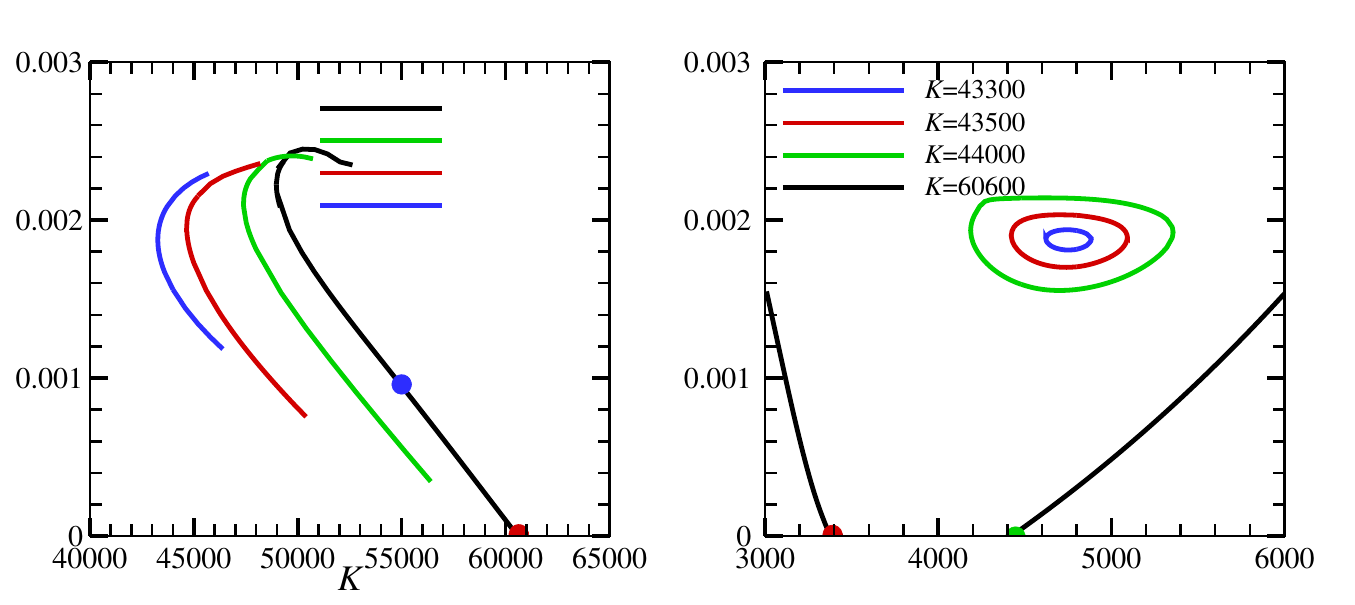}
\put(-4,39.3){(a)}
\put(33,36.4){$ \alpha_0=3380$}
\put(33,34){$\alpha_0=3550$}
\put(33,31.6){$\alpha_0=4000$}
\put(33,29.2){$\alpha_0=4800$}
\put(-2,21){\begin{turn}{90}{$\Delta Q$}\end{turn}}
\put(47,39.3){(b)}
\put(76,1){$ \alpha_0$}
\put(48,21){\begin{turn}{90}{$\Delta Q$}\end{turn}}
\end{overpic}
\caption{ 
Bifurcations of the BRE type travelling wave solutions from the curvature mode. The solution branches are computed by
(\ref{deaneq}) with $R=10^6$. 
(a) The scaled wavenumber $\alpha_0=\alpha R$ is fixed. (b) The Dean number $K$ is fixed.
The red and green points are the same as those in figure 4. 
Resolution is checked using up to $(L,M,N)=(25,25,30)$. }
\label{fig:change_zDn_different_alphaR_upper}
\end{figure}

\begin{figure}
\centering
\begin{overpic}[width=0.99 \textwidth]{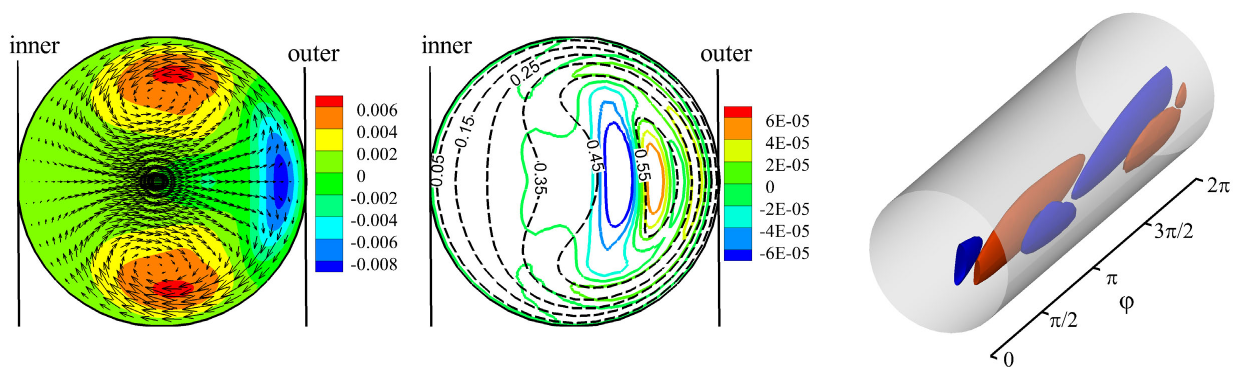}
\put(-4,24.5){(a)}
\put(30,24.5){(b)}
\put(66,24.5){(c)}
\end{overpic}
\caption{The same flow visualisation as figure 6, but for the BRE type solution at $(K,\alpha_0 ,R) =(5.5 \times 10^4, 3380, 10^6)$, corresponding to the blue point in  figure~\ref{fig:change_zDn_different_alphaR_upper}-(a)). 
The phase speed is $c\approx 0.5033$. In panel (c) the isosurfaces of $\tilde{\omega}_z=\pm 7 \times 10^{-5} $ are shown. 
}
\label{fig:BRE_vis}
\end{figure}

The equations (\ref{eq:BRE}) linearised around the Dean vortex is given by (\ref{eq:long}). Therefore, nonlinear travelling wave solutions bifurcating from the 2-vortex can be calculated from the neutral curve of the curvature mode seen in figure~\ref{fig:neutral}. As seen in figure~\ref{fig:change_zDn_different_alphaR_upper}, the bifurcation is subcritical. The parameter range for which we calculated nonlinear solutions is indicated by the dashed lines in figure~\ref{fig:neutral}; these solutions exist when $K$ is greater than 43086. Comparing figures~\ref{fig:Dn=750000_change_zBn_R=100000_VWI} and~\ref{fig:change_zDn_different_alphaR_upper}, the BRE type solutions, in contrast to the VWI type solutions, show an increase in flux relative to the 2-vortex. This is because, as seen in figure~\ref{fig:BRE_vis}-(a), the nonlinear interaction generates high-speed streaks near the upper and lower pipe walls. The three-dimensional wave components are concentrated around the outer side of the curved pipe (panel (c)). From panel (b), it can be observed that this location corresponds to the region where the streamwise velocity reaches its maximum.
Despite the qualitative differences in the flow fields compared to the VWI-type, the BRE solution also satisfies the shift-reflection symmetry (\ref{shiftre2}).

\section{Continuation from the finite-amplitude solutions in a straight pipe} \label{sec:Conti_pipe}

\subsection{Large Reynolds number limits of the exact coherent structures in a straight pipe} \label{sec:large_R_ECS}

As already noted, when $K=0$, the equations (\ref{eq:Dean_effect}) reduce to the straight pipe flow problem governed by the Navier-Stokes equations, for which a variety of exact coherent structures is available \citep{faisst2003,wedin2004,PK07,pringle2009,ozcakir2016} confirmed that some of these exact coherent structures follow the vortex-wave interaction theory at sufficiently high Reynolds numbers. In this paper, we utilise the solution found by \cite{PK07}, which is later labeled as M1 in \cite{pringle2009}. This solution was not studied in \cite{ozcakir2016}.

\begin{figure}
\centering
\begin{overpic}[width=0.9 \textwidth]{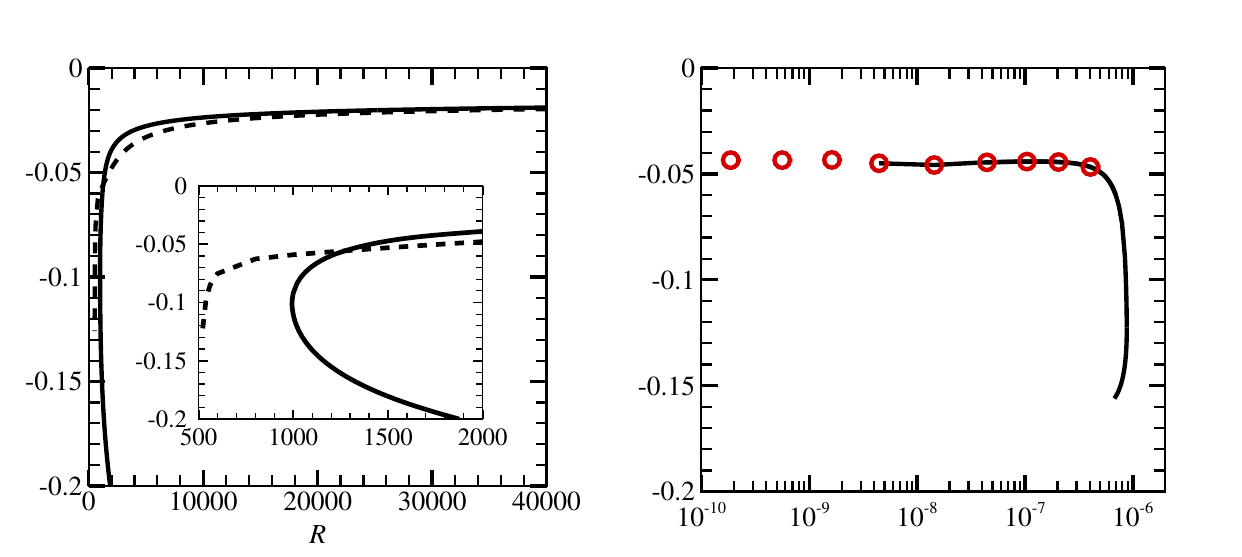}
\put(-2,21){\begin{turn}{90}{$\Delta Q$}\end{turn}}
\put(76,1){$\epsilon$}
\end{overpic}
\caption{ 
Continuation of the M1 straight pipe flow solution found by \cite{PK07}. (a) Results with a fixed wavenumber $\alpha=1.44$. The solid curve is the solution of the Navier-Stokes equations with $(L,M,N)=(70,50,6)$. 
The dashed curve shows the regularised VWI results with $(L,M)=(70,50)$. 
(b) Results with a fixed scaled wavenumber $ \alpha_0=\alpha R=1728$. The horizontal axis is  $\epsilon=R^{-2}$. The black curve and the red points correspond to the resolution levels $(L,M,N)=(50,22,24)$ and $(30,22,18)$, respectively. 
}
\label{fig:PK07_VWI_compare}
\end{figure}

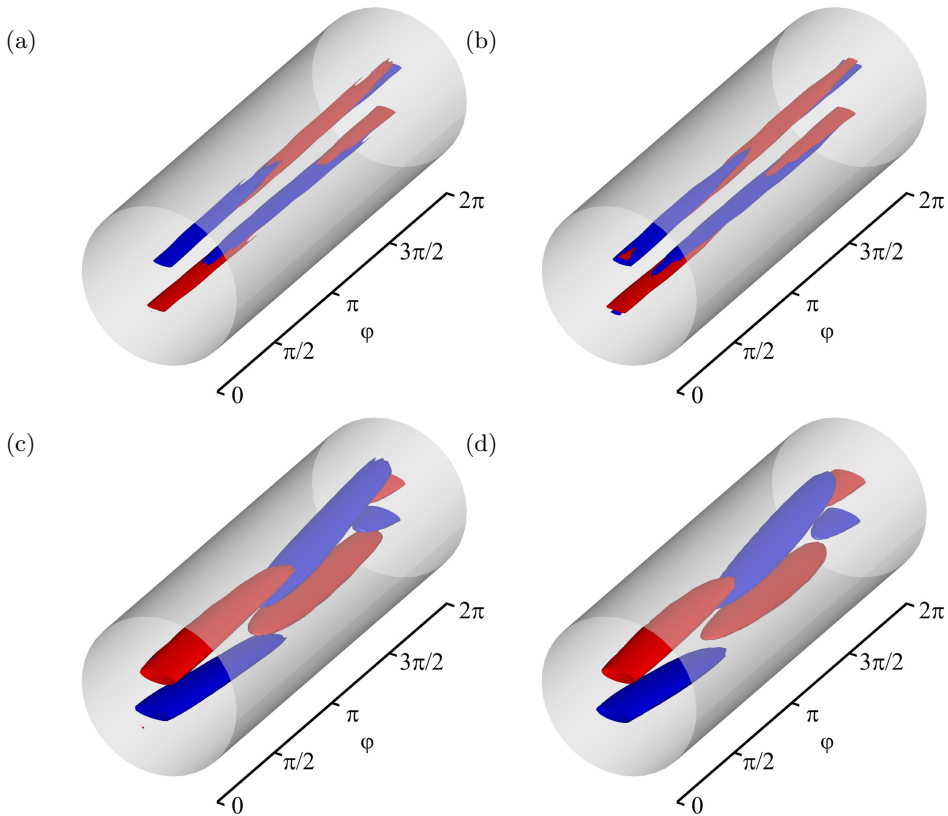
\begin{figure}
\centering
\begin{overpic}[width=0.9 \textwidth]{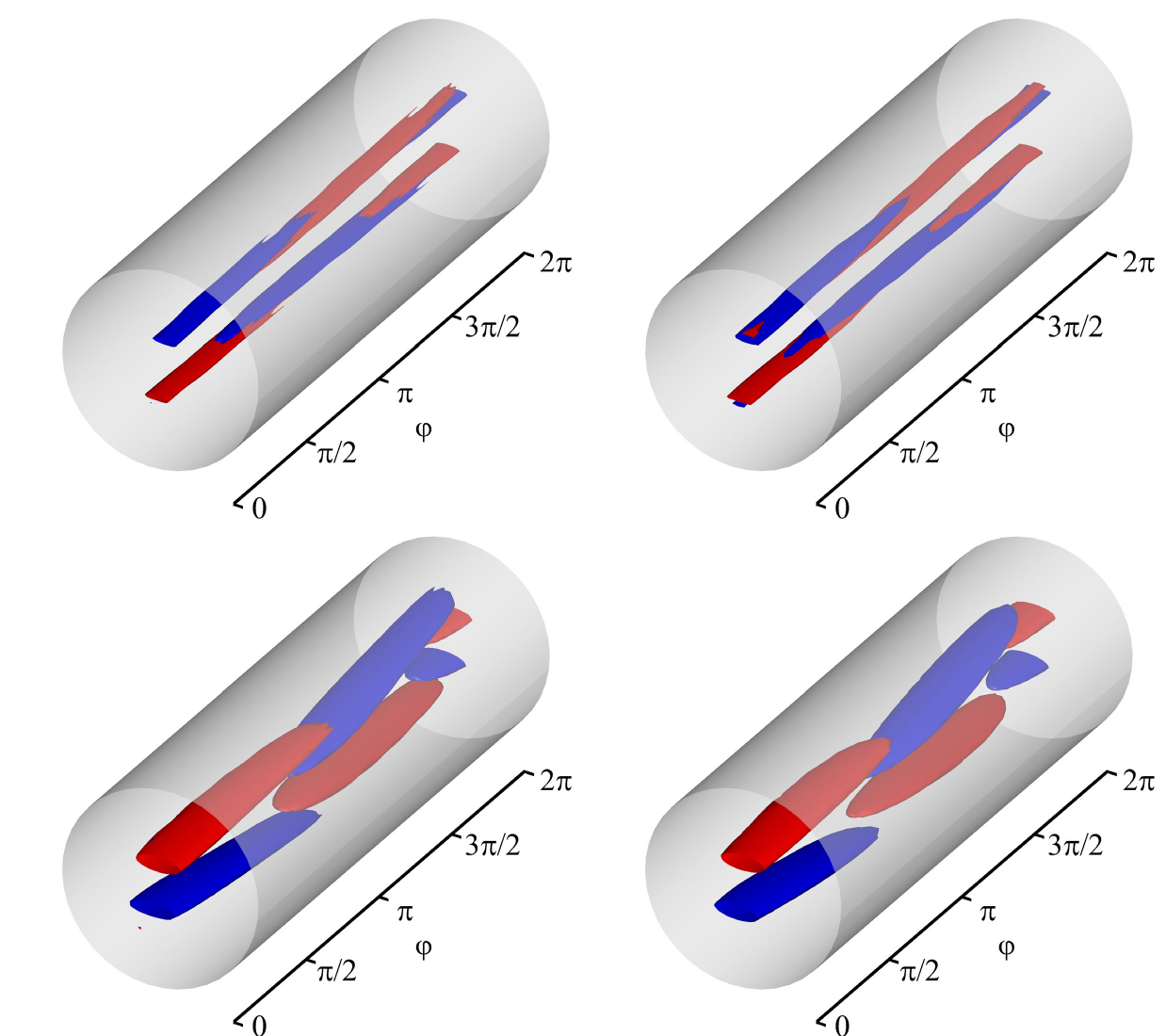}
\put(-3,39.3){(c)}
\put(47,83){(b)}
\put(-3,83){(a)}
\put(47,39.3){(d)}
\end{overpic}
\caption{The three-dimensional structure of the M1 solutions at the VWI- and BRE-limits. The same format as figure~\ref{fig:VWI_vis}-(c), but for $\tilde{\omega}_z= \pm 0.02$.
%
%
(a) The Navier-Stokes result at $(\alpha, R)=(1.44,4 \times 10^4)$.
(b) The regularised VWI result at $(\alpha, R)=(1.44,4 \times 10^4)$.
(c) The Navier-Stokes result at $( \alpha_0, R)=(1728, 10^4)$.
(d) The BRE result at $ \alpha_0=1728$ (i.e. formally $R=\infty$).
The isosurfaces are the asymptotic prediction at $R=10^4$, showing
the wave part of $10^{-4}r^{-1}(\frac{\partial (rV)}{\partial r}-\frac{\partial U}{\partial \theta})$.
}
\label{fig:VWI_BRE_Iso}
\end{figure}

The solid curve in figure \ref{fig:PK07_VWI_compare}-(a) shows the bifurcation diagram of the M1 solution. This solution appears at a saddle-node bifurcation, occurring at the lowest flux-based Reynolds number $2RQ\approx 773$ 
among all known solutions.
In the laminar parabolic profile, the value of $Q$ is 0.5, and under constant pressure, nonlinear effects should decrease the flux. Therefore, the upper curve in the figure corresponds to the `lower branch solutions' referred to in previous literature. As shown in \cite{PK07}, the solution possesses the mirror symmetry with respect to the line $\theta=\pm \pi/2$,
\begin{equation}
    [u,v,w](r,\theta,\varphi)=[u,-v,w](r,-\theta+\pi,\varphi),
\end{equation}
and the shift-reflection symmetry with respect to the line $\theta=0,\pi$,
\begin{equation}\label{shiftre}
    [u,v,w](r,\theta,\varphi)=[u,-v,w](r,-\theta,\varphi+\pi).
\end{equation}
Due to the symmetry of the system, arbitrary shifts in the $\theta$ and $z$ directions do not disqualify the M1 as a solution. However, when the effect of pipe curvature is introduced, the orientation in the $\theta$ direction is no longer arbitrary. In this study, we use both the original orientation from \cite{PK07} and an orientation rotated by 90 degrees in the $\theta$ direction. Figure~\ref{fig:VWI_BRE_Iso}-(a) shows the flow field for the latter orientation at $R=40000$, where
mirror-symmetry and shift-reflection symmetry become
\begin{equation}\label{mirror2}
    [u,v,w](r,\theta,\varphi)=[u,-v,w](r,-\theta,\varphi),
\end{equation}
and (\ref{shiftre2}), respectively.
Since the wavenumber $\alpha$ is fixed in figure~\ref{fig:PK07_VWI_compare}-(a), the solution branch is expected to converge to the VWI results at high Reynolds numbers. To verify this, we used the Navier-Stokes solution at $R=4\times 10^4$ as the initial condition for the regularised VWI code. The converged solution, shown in figure~\ref{fig:VWI_BRE_Iso}-(b), is almost indistinguishable from the initial condition, figure~\ref{fig:VWI_BRE_Iso}-(a). The dashed curve in figure~\ref{fig:PK07_VWI_compare}-(a) represents the regularised VWI results, which provide an excellent approximation when $R$ is $O(10^4)$ or larger. 


We can also compute the BRE limit of the M1 solution following \cite{Deguchi_Hall_Walton_2013}, where this limit was first applied for exact coherent structures in the context of plane Couette flow. The main idea is switching the parameters $(R,\alpha)$ to $({\alpha_0},\epsilon)$, where $\epsilon=R^{-2}$, and reformulating the problem as a regular perturbation problem. The limit as $\epsilon\rightarrow 0$ then corresponds to the BRE; see Appendix for more detail. 

The solid curve in figure~\ref{fig:PK07_VWI_compare}-(b) represents the same M1 solution as figure~\ref{fig:PK07_VWI_compare}-(a), but with ${\alpha_0}$ fixed and $\epsilon$ reduced. This computation, which employs a resolution deemed more than sufficient, encounters numerical instability, with the condition number of the Jacobian matrix deteriorating rapidly as $\epsilon$ decreases. This issue is somewhat expected, given that the limit involves an infinite Reynolds number and and infinitely long pipe. Lowering the resolution mitigates this issue (see the red circles in figure~\ref{fig:PK07_VWI_compare}-(b)), allowing the solution to even reach at $\epsilon=0$. Panel (c) of figure~\ref{fig:VWI_BRE_Iso} shows the high-resolution computation at $\epsilon=10^{-8}$ ($R=10^4$), while panel (d) presents the asymptotic prediction made at $R=10^4$ using the low-resolution result at $\epsilon=0$. Both results match closely, demonstrating that the use of $\epsilon=10^{-8}$ serves as a sufficiently accurate approximation of the BRE solution. 


\subsection{Continuation from the VWI mode} \label{sec:conti_VWI}

\begin{figure}
\centering
\begin{overpic}[width=0.96 \textwidth]{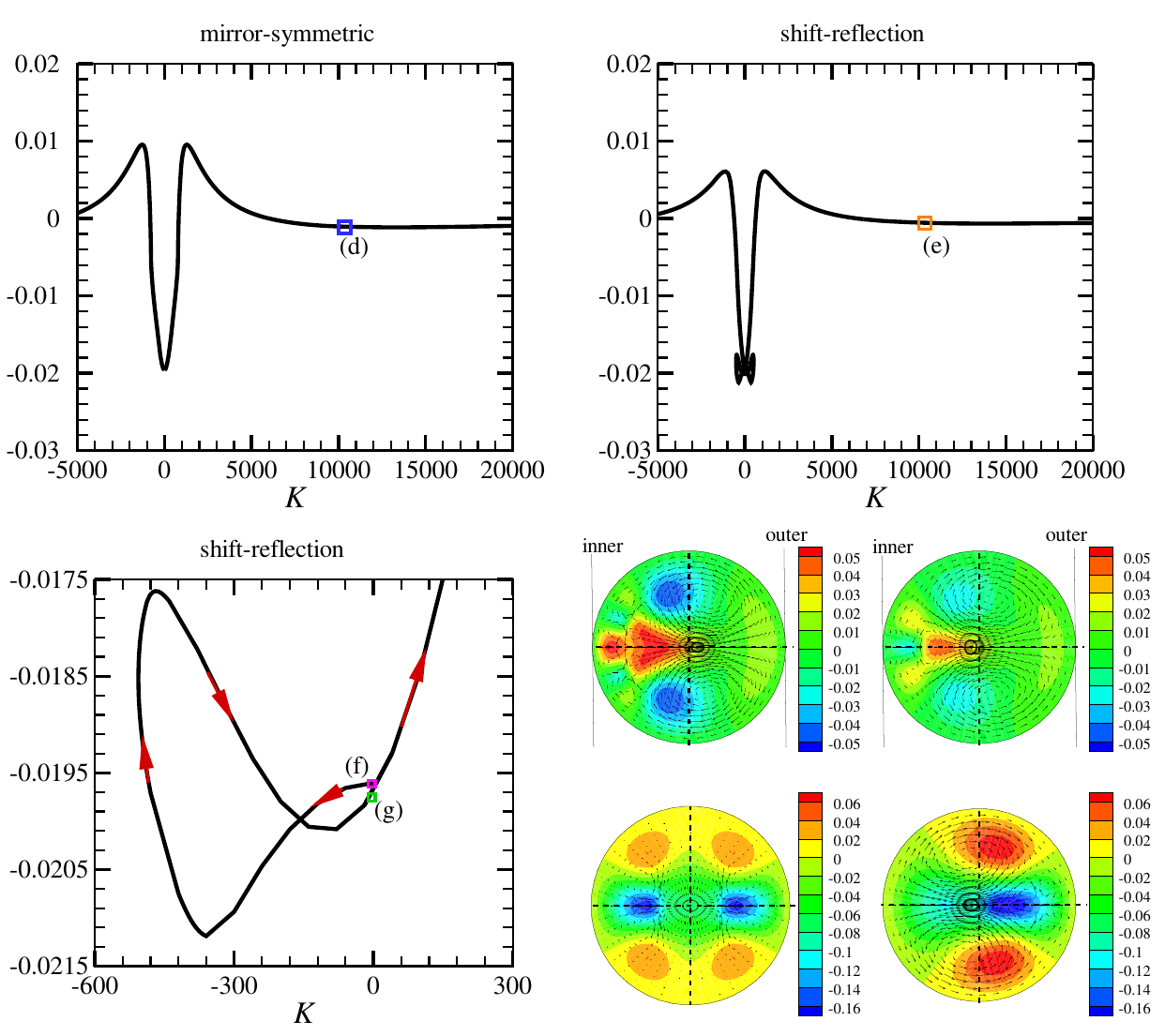}
\put(-2,21){\begin{turn}{90}{$\Delta Q$}\end{turn}}
\put(-2,65){\begin{turn}{90}{$\Delta Q$}\end{turn}}
\put(-3,43.1){(c)}
\put(47,83){(b)}
\put(-3,83){(a)}
\put(47,43.1){(d)}
\put(47,20){(f)}
\put(74,20){(g)}
\put(74,43.1){(e)}
\end{overpic}
\caption{
Continuation from the VWI type straight pipe flow solution (see figure~\ref{fig:PK07_VWI_compare}-(a)). The regularised VWI with $(\alpha,R)=(1.44,4\times 10^4)$ is used. Resolution is checked using up to $(L,M)=(70,50)$.
(a) The continuation from the rotated orientation shown in figure~\ref{fig:VWI_BRE_Iso}-(b). The solution has mirror-symmetry (\ref{shiftre}).
(b) The continuation from the original orientation. The solution has shift-reflection symmetry (\ref{mirror2}). (c) The same result as panel (b), but enlarged around $K=0$.
(d-g) Flow visualisation at the corresponding points on the bifurcation diagrams. The format is the same as figure~\ref{fig:VWI_vis}-(a).}
\label{fig:VWI_nBrA_sBrA}
\end{figure}
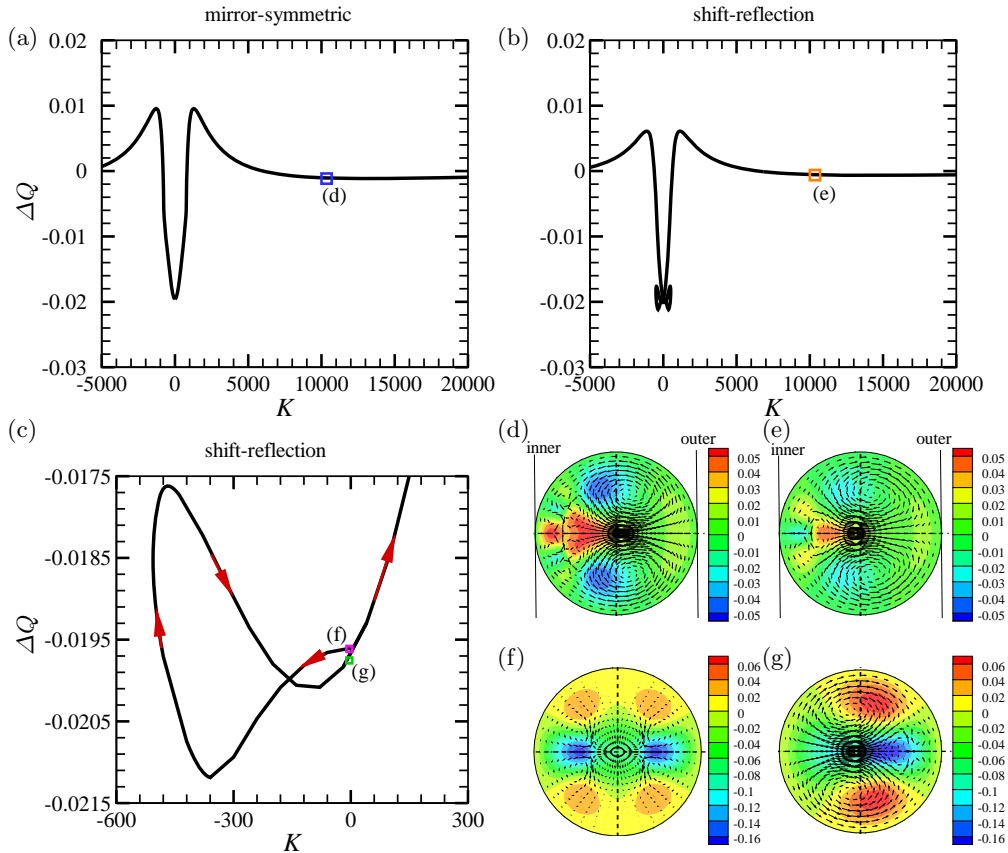
Now, let us add the effect of pipe curvature to the VWI mode obtained in figure~\ref{fig:PK07_VWI_compare}-(a). Starting from the configuration shown in figure~\ref{fig:VWI_BRE_Iso}-(b) (i.e. rotated by 90 degrees from the original orientation), mirror symmetry is preserved, while shift-reflection symmetry is broken. As shown in figure~\ref{fig:VWI_nBrA_sBrA}-(a), introducing a small curvature causes the value of $\Delta Q$ to increase.
As $K$ increases further, $\Delta Q$ decreases and approaches zero. We can extend the branch to values of $K$ where the linear instabilities are observed in section \ref{sec:LST_2v}. However, no connection to the 2-vortex solution was detected. 

The black dashed contours in figure~\ref{fig:VWI_nBrA_contour}-(a) illustrate the streak component $\overline{w}$ at $K=10000$ (indicated by the blue square in figure~\ref{fig:VWI_nBrA_sBrA}-(a)). The flow structure is overall similar to that for the 2-vortex solution at the same parameter (figure~\ref{fig:VWI_nBrA_contour}-(c)). The difference between those two fields, presented in figure~\ref{fig:VWI_nBrA_sBrA}-(d), reveals that changes in the streak due to three-dimensional effects primarily occur near the inner wall, contrasting with the observations in figures~\ref{fig:VWI_vis} and \ref{fig:BRE_vis}. The amplitude of the wave component responsible for this mechanism is strongest at the critical layer, as expected for VWI type exact coherent structures (see the coloured contours in figure~\ref{fig:VWI_nBrA_contour}-(a)).

If the original orientation (see figure~\ref{fig:VWI_nBrA_sBrA}-(f)) is used as the starting point of the continuation, the solution retains shift-reflection symmetry, but mirror-symmetry is lost. The resulting bifurcation diagram, shown in figure~\ref{fig:VWI_nBrA_sBrA}-(b), is similar to the previous one, except for the complex structure observed near $K=0$. figure~\ref{fig:VWI_nBrA_sBrA}-(c) is the enlargement of this part, where one of the symmetric branches has been omitted for clarity. Starting from the M1 solution, the solution branch extending into the negative $K$ region forms a loop and reaches the $K=0$ axis again. At this point, the solution retains shift-reflection symmetry but lacks mirror symmetry. The flow field of this solution (figure~\ref{fig:VWI_nBrA_sBrA}-(g)) closely resembles that of the asymmetric solution reported by \cite{PK07}, referred to as S1 in \cite{pringle2009}. 

The solution branch appears to continue indefinitely towards large $K$, but once again, no direct connection to Dean vortices is observed. The deviation of the streamwise velocity from the 2-vortex at $K=10000$, shown in figure~\ref{fig:VWI_nBrA_sBrA}-(e), is qualitatively similar to that in figure~\ref{fig:VWI_nBrA_sBrA}-(d). A strong vortex layer  appears in the wave component at the critical level, as shown in figure~\ref{fig:VWI_nBrA_contour}-(b). Although the structural details of the coloured contour differ from that in figure~\ref{fig:VWI_nBrA_contour}-(a), the physical role waves play seems to be the same.

\begin{figure}
\centering
\begin{overpic}[width=0.99 \textwidth]{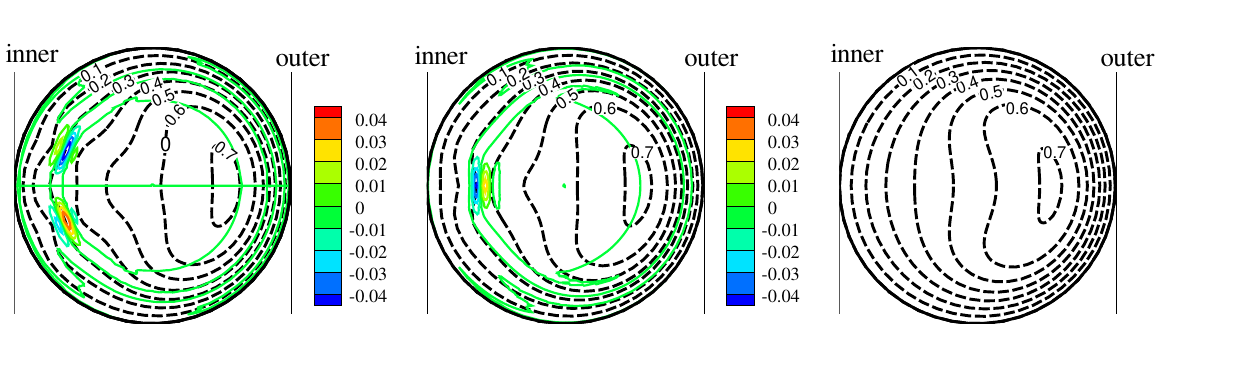}
\put(-5,24.5){(a)}
\put(30,24.5){(b)}
\put(62,24.5){(c)}
\end{overpic}
\caption{Contours of the streak (black dashed) and the wave vorticity (colourd solid). The same format as figure~\ref{fig:VWI_vis}-(b). $K=10^4$.
(a) The mirror-symmetric solution shown in figure~\ref{fig:VWI_nBrA_sBrA}-(d). The phase speed is $c\approx 0.3031$.
(b) The shift-reflection symmetric solution shown in figure~\ref{fig:VWI_nBrA_sBrA}-(e). The phase speed is $c\approx 0.3302$.
(c) The 2-vortex solution.
%
%
}
\label{fig:VWI_nBrA_contour}
\end{figure}

\subsection{Continuation from the BRE mode}\label{sec:conti_BRE}

\begin{figure}
\centering
\begin{overpic}[width=0.99 \textwidth]{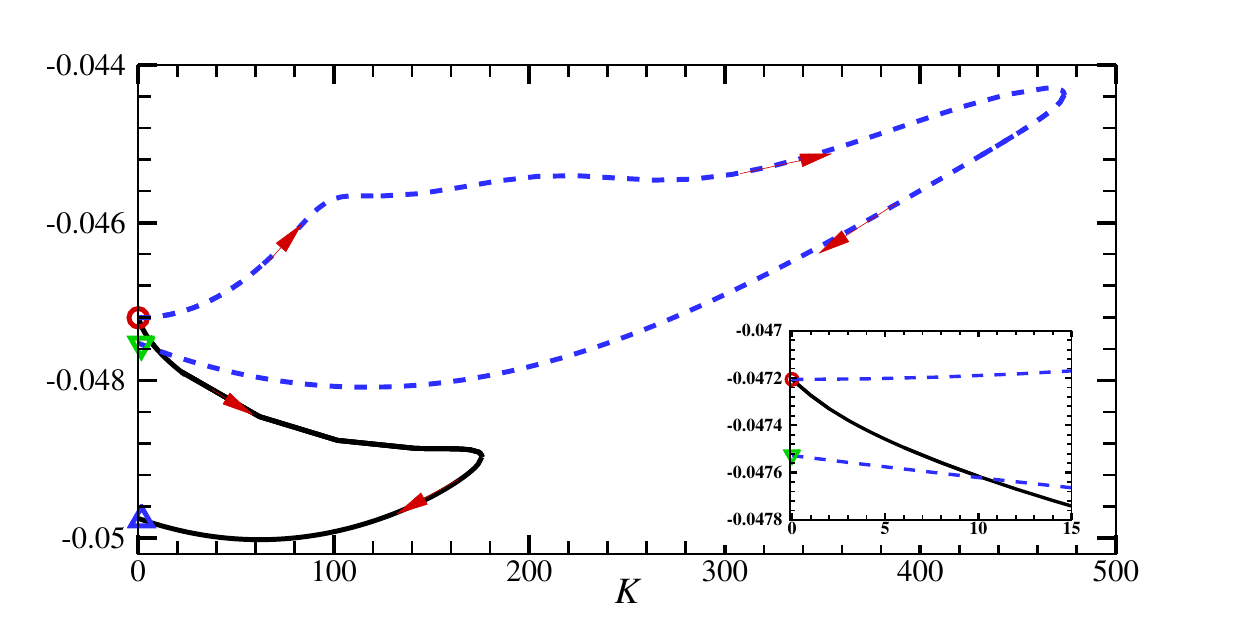}
\put(4,22){\begin{turn}{90}{$\Delta Q$}\end{turn}}
\end{overpic}
\caption{
Continuation from the BRE type straight pipe flow solution (see figure~\ref{fig:PK07_VWI_compare}-(b)). The equations (\ref{eq:Dean_effect}) are used with $( \alpha_0,R)=(1728,10^4)$. Resolution is checked using up to $(L,M,N)=(50,22,24)$. 
The continuation starts from the M1 solution indicated by the circle. 
The solid black curve shows the continuation from the rotated orientation shown in figure~\ref{fig:VWI_BRE_Iso}-(c). Along the curve mirror-symmetry is preserved.
The dashed blue curve illustrates the continuation from the original orientation. 
Along the curve shift-reflection symmetry is preserved. 
}
\label{fig:nBrA_zBn_R}
\end{figure}
\begin{figure}
\centering
\begin{overpic}[width=0.99 \textwidth]{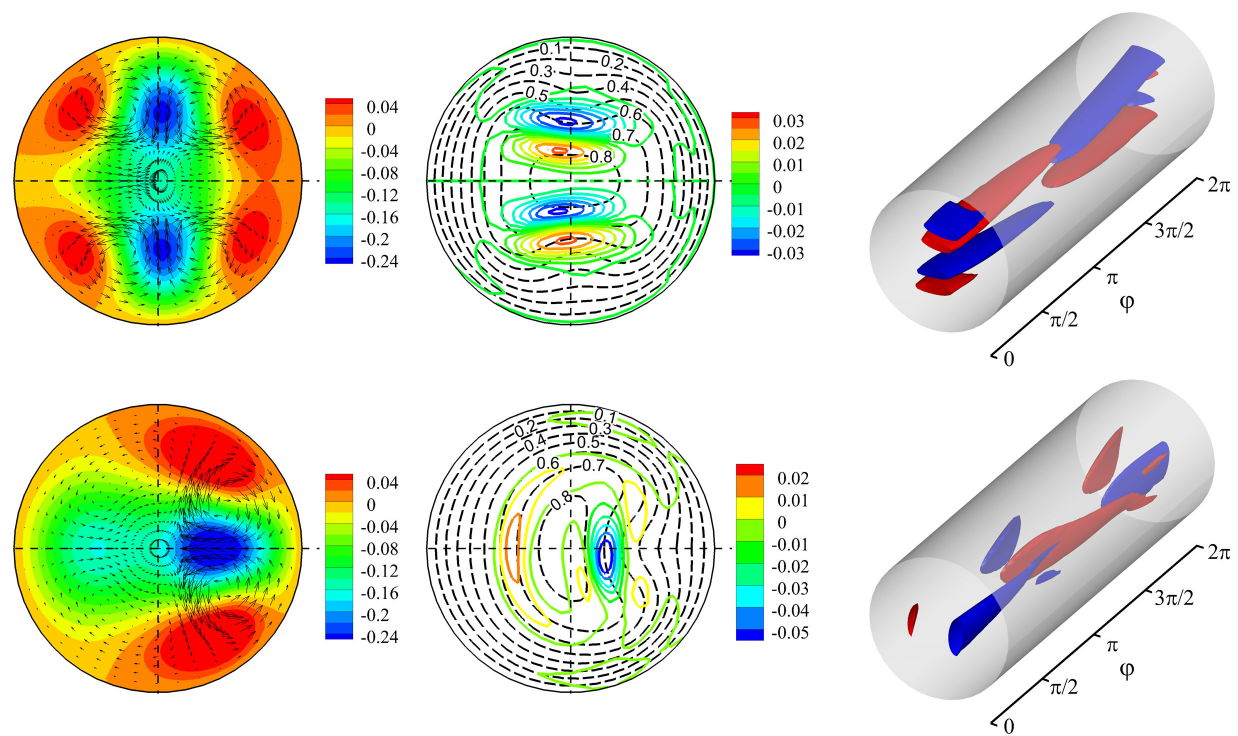}
\put(-3,24.5){(d)}
\put(32,24.5){(e)}
\put(66,24.5){(f)}
\put(-3,55){(a)}
\put(32,55){(b)}
\put(66,55){(c)}
\end{overpic}
\caption{
The asymmetric solutions obtained at $K=0$ in figure~\ref{fig:nBrA_zBn_R}. The same format as figure~\ref{fig:VWI_vis}. (c) and (f) illustrate isosurface at $|\tilde{\omega}_z|= 0.02$
(a-c) The solution at the upward triangle. The phase speed is $c\approx 0.6725$. (d-f) The solution at the downward triangle. The phase speed is $c\approx 0.6938$. 
%
}
\label{fig:sBrA_SB}
\end{figure}
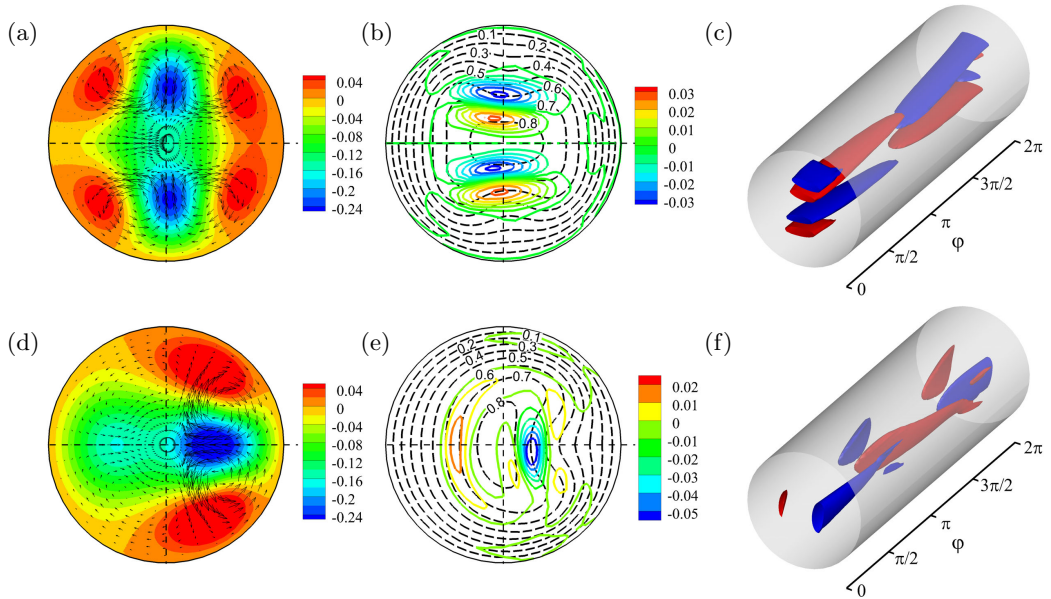

For completeness, we also examine the effect of curvature on the BRE-type solution. 
The black curve in figure~\ref{fig:nBrA_zBn_R} represents the bifurcation diagram starting from the M1 solution rotated by 90 degrees (indicated by the circle).
Along the solution branch, similar to figure~\ref{fig:VWI_nBrA_sBrA}-(a) for the VWI computation, mirror symmetry with respect to the $\phi=0,\pi$ axis is preserved. However, unlike the VWI case, a turning point is reached at relatively small $K$, after which the solution branch returns to $K=0$ (upward triangle). The straight pipe flow solution found at this point is not M1 but belongs to a previously unreported class of solutions that retain mirror-symmetry but lack shift-reflection symmetry. Nevertheless, as shown in figure~\ref{fig:sBrA_SB}-(a-c), the overall flow structure is not significantly different from M1. It is possible to continue the solution branch further from the new solution; however, the behavior of the solution branch is somewhat complicated, so it is not shown in the figure.

The blue curve in  figure~\ref{fig:nBrA_zBn_R} represents a similar computation, but initiated from the original orientation of M1. Throughout the computation, the shift-reflection symmetry is preserved. Again, the solution branch returns to the straight pipe (downward triangle). As shown in figure~\ref{fig:sBrA_SB}-(d-f), the flow field at this point closely resembles the S1 solution. While the solution branch can be further continued, it is unlikely that it can be extended for large $K$.

\section{Conclusion and discussions} \label{sec:conclusion}
In this study, we revisited the high Reynolds number limit in a weakly curved pipe, as originally considered by \cite{Dean_1927,Dean_1928}, adopting the same definition of the Dean number $K$. 
Dean and many subsequent studies employed the nonlinear equations~(\ref{deaneq}), which is independent of $z$ and $t$. We extended this result to account for three-dimensional travelling waves propagating downstream. 
Our approach is based on rational asymptotic analysis of the governing equations (\ref{fulleq}). The appropriate leading-order problem depends on how the axial wavenumber, $\alpha$, is scaled. We identified at least two distinct asymptotic states based on the suitable choices of this scaling.

If $\alpha$ is fixed while taking the limit, the leading-order problem can be formulated by combining the VWI theory with the Dean problem. More specifically, the rigorous asymptotic limit of this type is described by a system of equations that couples Dean's equation~(\ref{deaneq}) with a Rayleigh equation for the streak component, through the jump conditions (\ref{jumps}). However, from a computational perspective, it is more practical to use a system known as the regularised VWI, which replaces the Rayleigh equation and the jump conditions with the Orr-Sommerfeld equation and the Reynolds stress terms (\ref{Rstress}). The justification for using the regularised VWI is that, it serves as the Asymptotic Preserved Reduction (APR) of VWI, in the sense defined in section \ref{sec:numerics}.

If the product of $\alpha$ and the Reynolds number is fixed during the limiting process, the leading order system is given by (\ref{eq:BRE}), which represents the BRE with additional curvature terms included. 
However, numerical instability present in this limit, especially at high resolution, makes it preferable to use the augmented system (\ref{eq:Dean_effect}) at small $\epsilon=R^{-2}$ for computations. The validity of this approach is supported by the fact that system (\ref{eq:Dean_effect}) is the APR of (\ref{eq:BRE}). Moreover, since system (\ref{eq:Dean_effect}) retains all the necessary terms for the regularised VWI, it also serves as the APR of VWI. In this study, we used either system (\ref{eq:Dean_effect}) or the regularised VWI for numerical computations.

In section \ref{sec:BF_2v}, we obtained nonlinear travelling wave solutions through bifurcation analysis from the Dean vortices. This approach begins with the linear stability analysis of the 2-vortex solution (section \ref{sec:LST_2v}), which corresponds to the asymptotic limit of the neutral curve obtained by \cite{Canton_Schlatter_orlu_2016}. The instability detected by the Orr-Sommerfeld equation has an inviscid limit, from which VWI-type solutions undergo supercritical bifurcation. Given the flow characteristics and the nature of the bifurcation, this mode is likely part of the same family of supercritical travelling wave solutions identified by \cite{canton2020} at moderate curvature.

We also identified another linear instability, termed the curvature mode. This mode has long wavelengths and cannot be detected by the Orr-Sommerfeld equation. As the Dean number $K$ increases from zero, the curvature mode emerges first at $K\approx 5.72 \times 10^4$, followed by the onset of instability detected by the Orr-Sommerfeld equation at $K\approx 7.78 \times 10^5$.
The former critical Dean number agrees well with the recent full Navier-Stokes results from \cite{LCR_2024} and this is only slightly higher than the value at which the first 4-vortex solution appear, $K\approx 5.71 \times 10^4$. The nonlinear solutions that bifurcate subcritically from the curvature mode are described by the BRE and exist when $K$ exceeds $4.33 \times 10^4$. 

The two types of travelling wave solutions mentioned above exhibit distinct physical characteristics. First, the VWI-type solution increases the flow rate, while the BRE-type solution reduces it. Second, the Strouhal number $St=2\alpha c/Q$ is $O(1)$ for the VWI-type solution, while for the BRE-type solution, it is much smaller, on the order of $O(1/R)$. 
The Strouhal number here is defined in the same way as in the review by \cite{Vester_2016}, and it is frequently used to quantitatively investigate the swirl-switching phenomenon of Dean vortices.
Lastly, we note the connection to recent efforts to extend the concept of the self-sustaining process to Taylor-Couette flow, which also involves the shear-Coriolis instability. In this context, the supercritical process proposed by \cite{dessup2018} aligns with our VWI-type solution, while the subcritical process identified by \cite{WaAyDe22} is more suitable for our BRE-type solution.

In section \ref{sec:Conti_pipe}, we introduced curvature effects to the M1 solution, originally found by \cite{PK07} for straight pipes (i.e. $K=0$). 
In section \ref{sec:large_R_ECS}, we confirmed that this solution has a high Reynolds number asymptotic limit of VWI (BRE) type when $\alpha$ ($\alpha_0=\alpha R$) is fixed. In sections \ref{sec:conti_VWI} and \ref{sec:conti_BRE}, the effect of pipe curvature on those limits is examined.
Computational results for non-zero $K$ depend on how the initial M1 solution at $K=0$ is shifted in the azimuthal direction. The M1 solution possesses both mirror symmetry and shift-reflection symmetry, but their axes are orthogonal to each other. 
Starting the calculations from the orientation used by \cite{PK07} and from a 90-degree rotated orientation leads to different symmetries being preserved. 

We found that, in general, the branches of the VWI-type solutions can be extended to large values of $K$, whereas those for the BRE-type do not exhibit the same behaviour. The VWI branch approaches the 2-vortex solution as $K$ increases. However, even when the branch enters the region where the 2-vortex solution becomes unstable, we could not identify a bifurcation point connecting the two solutions. An intensive parameter search (not shown in this paper) concludes that there is no apparent connection between the three-dimensional instability of the 2-vortex solution and the self-sustained solutions in the straight pipe. This is quite different from plane Couette flow studied by \cite{Nagata_1988,Nagata_1990}, where three-dimensional solution generated by adding system rotation can be continued back to the zero-rotation limit. We note that \cite{barnes2000} reported similar difficulties in their study of pipe flow when secondary flow is induced by system rotation.

When $K$ is smaller than $5.71\times 10^4$, the 2-vortex solution is linearly stable, and therefore the transition to turbulence must be subcritical. Interestingly, figures~\ref{fig:change_zDn_different_alphaR_upper} and \ref{fig:VWI_nBrA_sBrA} suggest that in this regime, emergence of certain finite amplitude travelling waves, which may support turbulent activity, result in a positive value of $\Delta Q$, i.e. under the same pressure, the flow rate achieved is higher than that of the laminar flow. This implies that the so-called sub-laminar drag reported in previous numerical computations and experiments (see  \cite{noorani2015}, \cite{Vester_2016} ) is indeed realised for some exact coherent structures.
That said, in all our computations, the normalised flux $Q$ is always less than 0.5, which is the value for the laminar parabolic profile in the straight pipe. This result is consistent with the earlier observations.





\backsection[Acknowledgements]{
This research was supported by the Australian Research Council Discovery Project DP230102188.
}

\backsection[Declaration of Interests]{
The authors report no conflict of interest.
}

\appendix
\section{The computation of the BRE limit}


Applying the transformations $X=R^{-1}z$, $T=R^{-1}t$, $(u,v,w)=(R^{-1}U,R^{-1}V,W)$, the equations (\ref{eq:Dean_effect}) become
\begin{subequations}
\begin{eqnarray}
 \frac{\partial W}{\partial X}+\frac{\partial U}{\partial r}+\frac{U}{r}+\frac{1}{r}\frac{\partial V}{\partial \theta}=0,~~~\\
\mathcal{D}U-r^{-1}V^2- \frac{K}{2} W^2 \cos \theta = -\frac{\partial P}{\partial r}+ \triangle U-r^{-2}U-2r^{-2}\frac{\partial V}{\partial \theta}, \\
\mathcal{D}V+r^{-1}UV+ \frac{K}{2} W^2 \sin \theta=-r^{-1}\frac{\partial P}{\partial \theta}+\triangle V -r^{-2}V+2r^{-2}\frac{\partial U}{\partial \theta}, \\
\mathcal{D}W=-\epsilon \frac{\partial P}{\partial X}+4+\triangle W,
\end{eqnarray}
\end{subequations}
where $\mathcal{D}=\partial_T+U\partial_r+r^{-1}V\partial_{\theta}+W\partial_X$ and $\triangle=\partial_r^2+r^{-1}\partial_r+r^{-2}\partial_{\theta}^2+\epsilon \partial_X^2$. 
The flow parameters become $\epsilon=R^{-2}$, $K$, and the rescaled wavenumber. At this stage, no approximations have been made.

Similar to \cite{Deguchi_Hall_Walton_2013} for the Cartesian case, the appropriate poloidal-toroidal decomposition for this problem can be found as
\begin{eqnarray}
U=-r^{-2}\frac{\partial^2\tilde{\phi}}{\partial \theta^2}-\epsilon \frac{\partial^2\tilde{\phi}}{\partial X^2}-\frac{\partial \overline{\phi}}{\partial X},\\
V=\frac{\partial^2 (r^{-1}\tilde{\phi})}{\partial r\partial \theta}+\frac{\partial \psi}{\partial X},\\
W=\mathcal{W}+\epsilon r^{-1}\frac{\partial^2 (r\tilde{\phi})}{\partial r\partial X}+r^{-1}\frac{\partial (r\overline{\phi})}{\partial r}-r^{-1}\frac{\partial \psi}{\partial \theta},
\end{eqnarray}
where we require
\begin{eqnarray}
\int^{2\pi}_0\overline{\phi}d\theta=\overline{\phi},\qquad
\int^{2\pi}_0 \tilde{\phi} d\theta=0.
\end{eqnarray}
We can compute the solutions for $\epsilon=0$ (i.e. the BRE limit) using those potentials.

\bibliographystyle{jfm}  
\bibliography{Reference}  

\end{document}